\documentclass[10pt,journal,compsoc]{IEEEtran}

\newfont{\mycrnotice}{ptmr8t at 7pt}
\newfont{\myconfname}{ptmri8t at 7pt}

\newcommand{\Comment}[1]{\relax}          
\newcommand{\cut}[1]{}

\usepackage{color}

\def\compactify{\itemsep=0pt \topsep=0pt \partopsep=0pt \parsep=0pt}
\let\latexusecounter=\usecounter
\newenvironment{CompactEnumerate}
 {\def\usecounter{\compactify\latexusecounter}
  \begin{enumerate}}
 {\end{enumerate}\let\usecounter=\latexusecounter}

\usepackage{latexsym}
\usepackage{times}
\usepackage{graphics}
\usepackage{multirow}
\usepackage{color}
\usepackage{amssymb,amsmath,amsfonts}
\usepackage{wrapfig,epsfig}
\usepackage[center]{caption}
\usepackage{subfigure}

\begin{document}
\begin{sloppypar}

\title{AIDE: An  Automated Sample-based Approach for Interactive Data Exploration }

\author{
Kyriaki Dimitriadou$^*$, Olga Papaemmanouil$^*$ and  Yanlei Diao$^\dagger$\\
{$^*$ Brandeis University, Waltham, MA, USA, $^{\dagger}$ University of Massachusetts, Amherst, MA, USA}\\
       
{$^*$\{kiki,olga\}@cs.brandeis.edu, $^{\dagger}$yanlei@cs.umass.edu}
}

\IEEEtitleabstractindextext{%
\begin{abstract}

In this paper, we argue that database systems be augmented with an automated data exploration service that  methodically steers users through the data in a meaningful way. Such an automated system is crucial for deriving insights from complex datasets found in many big data applications such as scientific and healthcare applications as well as for reducing the human effort of data exploration. Towards this end, we present  AIDE, an Automatic Interactive Data Exploration framework that assists users in discovering new interesting data patterns and eliminate expensive ad-hoc exploratory queries. 

AIDE relies on a seamless integration of classification algorithms and data management optimization techniques that collectively strive to accurately learn the user interests based on his relevance feedback on strategically collected samples. We present a number of exploration techniques as well as optimizations that minimize the number of samples presented to the user while offering interactive performance. {AIDE can deliver highly accurate query predictions for very common conjunctive queries with small user effort while, given a reasonable number of samples, it can predict with high accuracy  complex disjunctive queries. It provides interactive performance as it limits the user  wait time per iteration of exploration to less than a few seconds. \cut{Our user study also shows that AIDE  improves the current state-of-the-art of manual exploration by significantly reducing the user effort and total exploration time.}}

\end{abstract}

}

\maketitle


\vfill\eject
\section{Introduction}

Traditional data management systems assume that when users pose a query they a) have good knowledge of the schema, meaning and contents of the database and b) they are certain that this particular query is the one they wanted to pose. In short, traditional DBMSs are designed for applications in which the users know what they are looking for. \cut{The above assumption is valid for numerous past and present applications.} However, as  data are being collected and stored at an unprecedented rate, we are building more dynamic  data-driven applications where this assumption is not always true. \cut{Indeed, managing an employee or an inventory database is a drastically different setting than looking for interesting patterns over a scientific database}

\emph{Interactive data exploration} (IDE)  is one such example. In these applications, users are trying to  make sense of the underlying data space by  experimenting with queries, backtracking on the basis of query results and rewriting their queries aiming to discover interesting data objects.  IDE  often incorporates ``human-in-the-loop'' and it  is fundamentally  a long-running, multi-step process with the user's interests  specified in imprecise terms.  


%

One application of IDE  can be found in the domain of evidence-based medicine (EBM). Such  applications often involve the generation of systematic reviews, a comprehensive assessment of the totality of evidence that addresses a well-defined question, such as  the effect on mortality of giving versus not giving drug A  within three hours of a symptom B.  While a content expert can  judge whether a given clinical trial  is of interest or not (e.g., by reviewing parameter values  such as disease, patient age, etc.), he often does not have a priori knowledge of the exact attributes that should be used to formulate a query to collect all relevant clinical trials. Therefore the user relies on an ad hoc  process that includes three steps: 1) processing numerous selection queries with iteratively varying selection predicates, 2)  reviewing returned objects (i.e., trials) and classifying them to relevant and irrelevant, and 3) adjusting accordingly the selection query for the next iteration. The goal here is to discover the selection predicates that balances the trade-off between collecting all relevant objects and reducing the size of  returned results. These ``manual'' explorations  are  typically labor-intensive: they may take  days to weeks to complete since users need to examine thousands of  objects.

Scientific applications, such as ones analysing astrophysical surveys (e.g.,~\cite{LSST, SloanSky}), also suffer from similar situations. Consider an astronomer looking for interesting patterns over a scientific database: they do not know what they are looking for, they only wish to find interesting patterns; they will know that something is interesting only after they find it.  In this setting, there are no clear indications about how  the astronomers should formulate their queries.  Instead, they may want to navigate through a subspace of the data set (e.g., a region of the sky) to find  objects of interest, or may want to see a few samples, provide yes/no feedback, and expect the  system to find more similar objects. 
\cut{Other  IDE applications include but are not limited to financial analysis and genomics.}

To address the needs of  IDE applications, we propose an \emph{Automatic Interactive Data Exploration (AIDE)} framework that automatically discovers data relevant to her interest. Our approach unifies the three  IDE steps---query formulation, query processing and result reviewing---into a single automatic process, significantly reducing the user's exploration effort and the  overall exploration time.  In particular, an AIDE user engages in a ``conversation'' with the system indicating her interests, while in the background the system  builds a user model that predicts data matching these interests. 


AIDE offers an iterative exploration model: in each iteration the user is  prompted to provide her feedback on  a {set} of sample objects  as relevant or irrelevant to her exploration task. Based on her feedback,  AIDE generates the user's  exploration profile, i.e., a  user model that  classifies database objects as relevant or irrelevant.  AIDE  leverages this model to explore further the data space, identify strategic sampling areas and collect new samples for the next iteration. These  samples are   presented to the user  and her new  feedback is  incorporated into the user model. This iterative process aims to generate a user model that identifies all relevant objects  while eliminating  the misclassification of irrelevant ones.

%

AIDE's model raises new challenges. First, AIDE operates on the unlabeled space of the whole data space that the user aims to explore. To offer effective  exploration results (i.e., accurately predict the user's interests)  it has to decide and retrieve in an \emph{online} fashion the example objects to be extracted and labeled by the user.  Second, to achieve desirable interactive experience for the user,  AIDE  needs  not only to provide accurate results, but also to minimize the number of samples presented to the user (which determines the amount of user effort)  as well as  to reduce the sampling and space exploration overhead (which determines the user's wait time in each iteration). 

These challenges  cannot be addressed by existing machine learning techniques.  Classification algorithms (e.g.,~\cite{breinman84}) can be leveraged to build the user  model  and the information retrieval community offers solutions on incrementally incorporating relevance feedback in these models (e.g.,~\cite{zhou03}). However, these approaches operate under the assumption that the  sample set shown to the user  is either known a priori or, in the case of online classification, it is provided  incrementally by a different party. In other words, classification algorithms do not deal with  \emph{which} data samples to show to the user, which is one of the main research challenges for AIDE.

Active learning systems~\cite{Settles10activelearning} also extract unlabeled samples to be labeled by a user and the goal is to achieve high accuracy using as few labeled samples as possible, therefore minimizing the user's labeling effort. In particular, pool-based sampling techniques selectively draw samples from a large pool of unlabeled data. However, these solutions exhaustively examine \emph{all} unlabeled objects in the pool in order to identify the best samples to show to the user based on some informativeness measure~\cite{roy2001}. Therefore, they implicitly assume negligible sample acquisition costs and hence cannot offer  interactive performance on big data sets  as expected by IDE applications. In either case, model learning and sample acquisition are decoupled, with the active learning algorithms not addressing the challenge of \emph{how} to minimize the cost of sample acquisition.

To address the above challenges, AIDE closely \emph{integrates} classification model learning (from existing labeled samples) and effective data exploration and sample acquisition (deciding best data areas to sample). Specifically, our techniques leverage the classification properties of decision tree learning to identify promising data exploration areas from which new  samples are extracted, as well as to minimize the number of samples shown to the user. These techniques aim  to predict linear patterns of user interests, i.e., we assume relevant objects are clustered in multi-dimensional hyper-rectangles. These interests can be  expressed as range  queries with disjunctive and/or conjunctive predicates.  

This paper extends our previous our previous work on automatic data exploration~\cite{aide_sigmod14,aide_vldb15}. Specifically, we extended AIDE with a number of performance optimizations that are designed to reduce  the total exploration overhead. \cut{This overhead  includes  the number the number of samples labeled by the user, the convergence rate to an accurate user model as well as the user's wait time.} Specifically,  we introduce: (a) a skew-aware exploration technique that deals with both uniform and skewed data spaces as well as user interests that lie on either the sparse or dense parts of the distribution, (b) a probabilistic sampling strategy for selecting the most informative sample to present to the user; the strategy is designed to reduce the user's exploration effort and (c) an extended relevance feedback model that allows users to annotate ``similar'' (rather than only relevant/irrelevant) samples, allowing us to further reduce the total exploration time. We also include a new set of experimental results that demonstrate the effectiveness and efficiency of our new exploration techniques.
%

The specific  contributions of this work are the following: 

\begin{CompactEnumerate}
\item  We introduce AIDE, a novel,  automatic data exploration framework, that  navigates the user throughout the  data space he wishes to explore. AIDE relies on the user's feedback on example objects to generate a user model that predicts data relevant to the user. It employs a unique combination of machine learning, data exploration, and sample acquisition techniques to deliver highly accurate predictions of linear patterns of user interests with interactive performance. Our data exploration techniques  leverage the properties of classification models to  identify \emph{single} objects of interest, expand them to more accurate \emph{areas of interests},  and progressively refine the prediction of these areas. Our techniques address the trade-off between quality of results  (i.e., accuracy) and   efficiency (i.e., the total {\em exploration time} which includes the  total sample reviewing time and wait time by the user). 


\item We introduce new optimizations  that address the presence of skew in the underlying exploration space as well as a novel probabilistic approach for identifying the most informative sample set to show to the user. We also include an extended feedback model based on which the user can also indicate similar but not necessarily relevant objects. This new model allows  AIDE to  focus its exploration on  on certain promising domain ranges reducing significantly the user's labeling effort. 

\item We evaluated our implementation of AIDE using the SDSS database~\cite{SloanSky} and a user  study. { Our results indicate that AIDE and its novel optimizations are highly effective and efficient. AIDE can predict common conjunctive queries with a small number of samples, while given an acceptable  number of labeled samples it predicts highly complex disjunctive queries with high accuracy.   AIDE also offers interactive performance as the  user wait time per iteration is less than a few seconds in average. Our user study revealed that AIDE can reduce the user's labeling effort by up 87\%, with an average of 66\% reduction. When  including the sample reviewing time, it  reduced the total exploration time by 47\% in average.
}

\end{CompactEnumerate}

The rest of the paper is organized as follows. Section~\ref{s:system}  outlines the AIDE framework and Section~\ref{s:exploration_overview} describes  the  phases of our data exploration approach. Section~\ref{s:optimizations} discusses the new performance optimizations we introduce in AIDE.  Section~\ref{s:experiments} presents our experimental results.  Section~\ref{s:related} discusses the related work and we conclude in Section~\ref{s:conclusions}. 

\section{AIDE Framework Overview}\label{s:system}
In this section we introduce our system model, describe  the classification algorithms we use and provide a definition of our exploration problem. 

\begin{figure}
\centering 
 \includegraphics[totalheight=3.1in, angle=90] {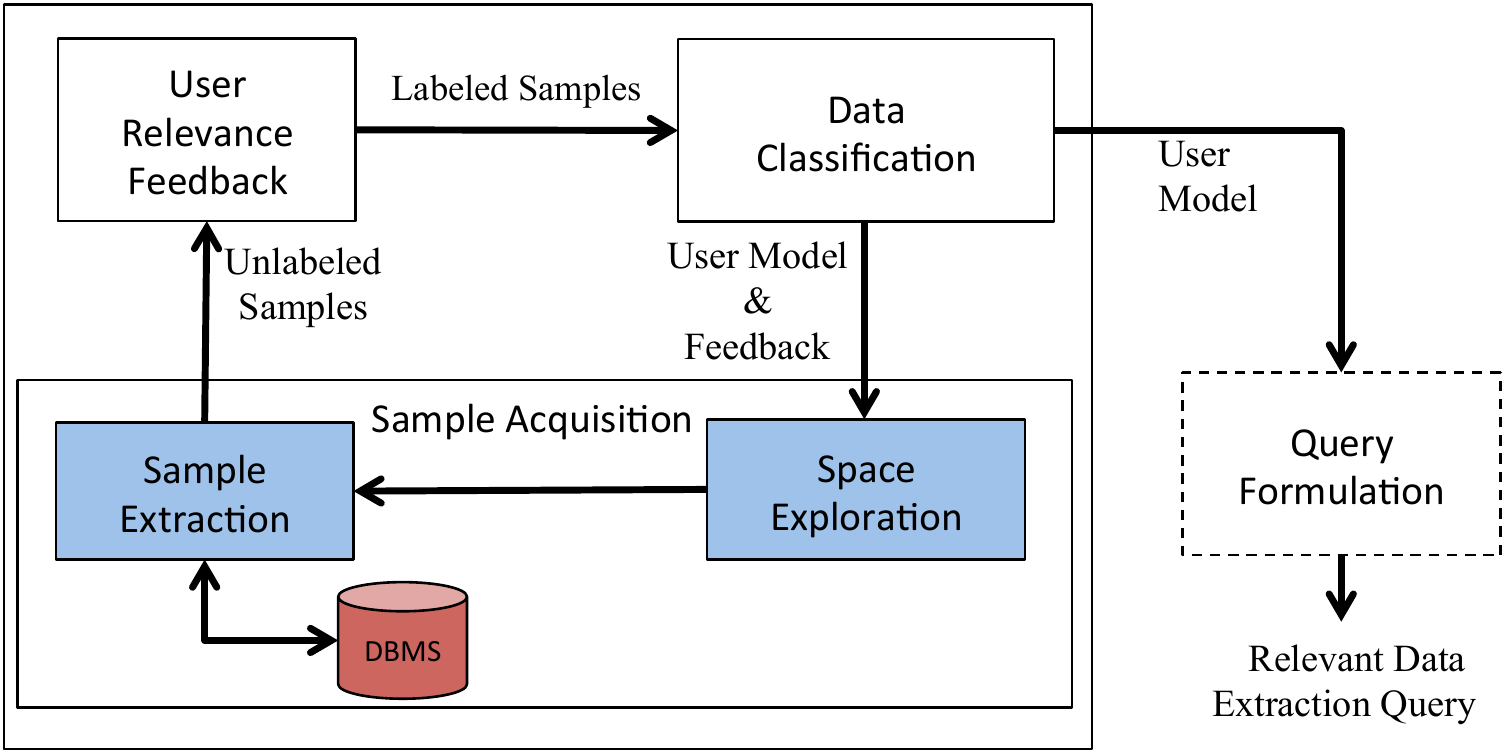}
\caption{\small{Automated Interactive Data Exploration Framework.}} \label{f:autoframework}
\vspace{-4mm}
\end{figure}

\subsection{System Model}\label{s:systemmodel}

The workflow of our exploration framework is depicted in Figure~\ref{f:autoframework}. AIDE presents to the user  sample database objects and requests her feedback on their relevance to her exploration task, i.e., characterize them as relevant or not.  For example, in the domain of evidence-based medicine, users are shown sample clinical trials  and they are asked to review their abstract and their  attributes (e.g., year, outcome, patience age, medication dosage, etc) and label each sample trial as interesting or not.  AIDE allows also the user to annotate samples that are similar (in some attribute) but not match exactly her interest, by marking them as ``similar'' samples. Finally, the user can modify her feedback on previously seen samples, however this could potentially prolong the exploration process.

The iterative steering process starts when the user provides her feedback by labeling samples are relevant or not. The relevant and irrelevant samples  are used to train a binary classification model  that  characterizes the user's interest,  e.g., it predicts which clinical trials are relevant to the user  based on the  feedback collected so far ({\em Data Classification}) \footnote{"Similar" samples  are not included in the training of the user model. In Section~\ref{s:maybe} we discuss in detail how we leverage these samples.}. This model may use any subset of the object's attributes to characterize  user interests.  However, domain experts could leverage their domain knowledge to restrict the attribute set on which the exploration is performed. For instance, one could request an exploration only on the attributes \textit{dosage} and \emph{age}. In this case,  relevant  trials will be characterized on a subset of these attributes (e.g.,  relevant trials have dosage $>$45mg).

In each iteration, more samples  (e.g., records of clinical trials) are extracted and presented to the user for feedback.  AIDE leverages the current user model as well as the user's feedback so far to  identify promising sampling areas ({\em Space Exploration}) and  retrieve the next sample set from the database (\emph{Sample Extraction}).  New labeled objects  are incorporated with the already labeled sample set and a new classification model is built. The  steering process is completed when the user terminates the process explicitly, e.g., when reaching a satisfactory set of relevant objects or when she does not wish to label more samples. Optionally, AIDE ``translates'' the  classification  model  into a query expression (\emph{Query Formulation}). This query  will retrieve  objects characterized as relevant by the user model ({\em Data Extraction Query}).

AIDE strives to converge to a model  that captures the user interest, i.e., eliminating irrelevant objects while identifying a large fraction of relevant ones. Each round refines the user model by exploring further the data space.  The user decides on the effort he is willing to invest (i.e., number of samples he labels) while AIDE leverages his feedback to strategically sample the exploration space, i.e., collect samples  that  improve the accuracy of the  classification model.  The more effort invested in this iterative process, the more effective the user model will be. \cut{However, higher numbers of samples increase the sample extraction and processing time, i.e., the user {\em wait time} between iterations.} 

\begin{figure}[t]
\centering 
 \includegraphics[totalheight=2.3in, angle=-90] {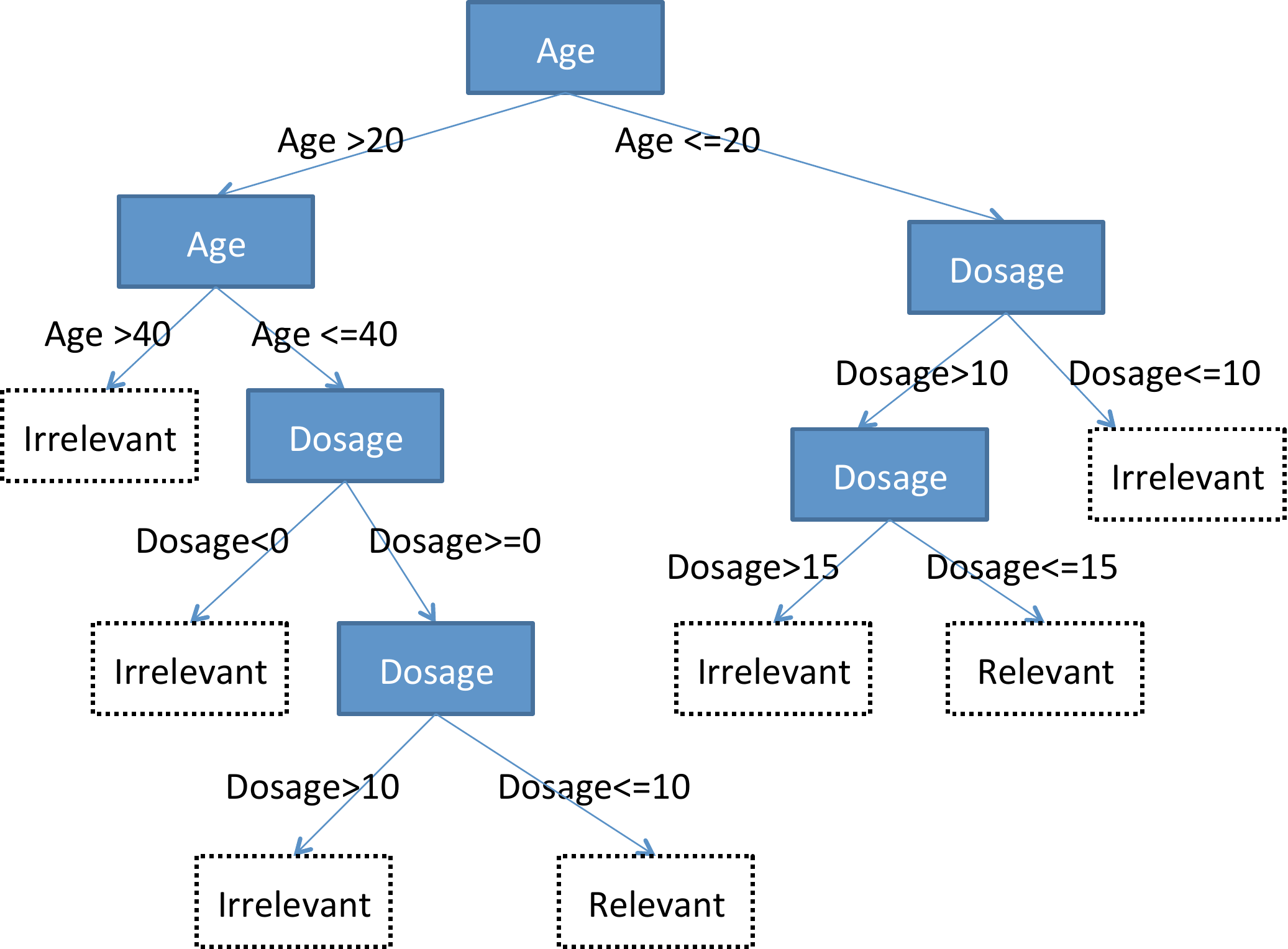}
\caption{{\small {An example decision tree.}}} 
\label{f:dtree}
\vspace{-4mm}
\end{figure}

\subsection{Data Classification \& Query Formulation}\label{s:system_classifier}
AIDE relies on decision tree classifiers to identify linear patterns of user interests, i.e., relevant objects clustered in multi-dimensional hyper-rectangles. Decision tree learning~\cite{breinman84}  produces classification models that predict the class of an unclassified object based on  labeled training data.  The major advantage of decision trees is that they provide easy to interpret  models that clearly describe the features characterizing each data class. Furthermore, they perform well with large data and the decision conditions of the model can be easily translated to simple boolean expressions. This  feature is  important since it allows us to map decision trees to queries that retrieve the relevant data objects. 

Finally,  decision trees can handle both numerical and categorical data. This allows  AIDE to operate on both data types assuming a  distance function is provided to calculate the similarity between two data objects. Measuring the similarity between two  objects is a requirement of the space exploration step. AIDE treats the similarity computation as an orthogonal step and can make use of any distance measure. For continuous data sets (e.g., numerical),  the Euclidean distance can be used. Computing similarity between categorical data is more challenging due to the fact that there is no specific ordering between categorical values. However, a number of similarity measures have been proposed in the literature for categorical data, and AIDE can be extended in a straightforward way to incorporate them.

{\bf Query Formulation} Let us assume a decision tree classifier that predicts relevant and irrelevant  clinical trials objects based on the attributes {\tt age} and {\tt dosage} (Figure~\ref{f:dtree}). This tree provides  predicates that characterize the relevant class and predicates that describe the irrelevant class.  In Figure~\ref{f:dtree},  the relevant class is described by the predicates $(age \le  20  \wedge 10 < dosage \le 15)$ and $(20< age \le 40 \wedge 0 \le dosage \le 10)$, while the irrelevant class is characterized by the predicates $(age \le 20 \wedge dosage \le 10)$ and $(20<age  \le 40 \wedge  dosage > 10)$ (here we ignore the predicates that refer to values outside attribute domains, such as $age>40$, $age<0$, $dosage<0$ and $dosage>15$). 
Given the decision tree in Figure~\ref{f:dtree} it is straightforward to formulate the extraction query for the relevant objects ({\it select  * from table where (age $\le$ 20 and dosage $>$10 and dosage $\le$ 15) or (age $>$ 20 and age $\le$ 40 and dosage $\ge$ 0 and dosage $\ge$ 10)}).

\subsection{Problem Definition}\label{s:problemdef}

Given a database schema $\mathcal{D}$, 
let us assume the user has decided to focus his exploration on $d$ attributes, where these  $d$ attributes may include both attributes relevant and those irrelevant to the final query that represents the true user interest.
Each exploration task is then performed in a $d$-dimensional space of $T$  tuples where each tuple represents an object characterized by $d$ attributes.   For a given user,  our exploration space is divided to the relevant object set $T^r$ and irrelevant set $T^{nr}$. Since the user's interests are unknown to AIDE, the sets $T^r$ and $T^{nr}$ are also unknown in advance. 

AIDE aims to generate a  model that predicts these two sets, i.e., classifies a tuple in $T$ as relevant or irrelevant. To achieve that, it iteratively trains a decision tree classifier. Specifically,   in each iteration $i$, a sample tuple set $S_i \subseteq T$ is shown to the user and his relevance feedback assigns these samples to two data classes, the relevant  object class $D^r \subseteq T^r$, and the irrelevant one, $D^{nr} \subseteq T^{nr}$.  Based on the samples assigned to these  classes up to the $i$-th iteration, a new decision tree classifier $C_i$ is generated. This classifier corresponds to a predicate set ${P_i^r \bigcup P_i^{nr} }$, where the predicates ${P_i^r}$ characterize the relevant class and predicates ${P_i^{nr}}$ describe the irrelevant one.

We measure AIDE's effectiveness (aka accuracy of a classification model)  by evaluating the $F$-measure, the harmonic mean between precision and recall.\footnote{Here, if $tp$ are the true positives results of the classifier (i.e., correct classifications as relevant), $fp$ are the false positives (i.e., irrelevant data classified as relevant) and $fn$ are the false negatives (i.e., relevant data classified as irrelevant), we define the precision  of our classifier  as $precision=\frac{tp}{ tp + fp}$   and the recall as $recall=\frac{tp}{tp+fn}$.} 
Our goal is to maximize the $F$-measure of the final decision tree $C$ on the total data space $T$,  defined as:
$F(T)=\frac{2 \times precision(T) \times recall(T)}{precision(T)+recall(T)}
$.
The perfect precision value of 1.0 means that every object characterized as relevant by the decision tree is indeed relevant,  while a good recall ensures that our final query can retrieve a good percentage of the relevant to the user objects. 


\section{Space Exploration Techniques}\label{s:exploration_overview}
Our main research focus is on optimizing the effectiveness of the exploration (i.e., the accuracy of the final user model)  while offering interactive experience to the user. To address that AIDE strives to improve on a number of efficiency factors, including the number of samples presented to the user  and the number of sample extraction queries processed in the backend. In this section, we introduce our main exploration techniques that tackle these goals. 

AIDE assumes that  user interests are captured by \emph{range queries}, i.e., relevant objects are clustered in one or more areas in the data space. Therefore, our goal is to generate a user model that predicts \emph{relevant areas}. The user model can then be translated to a range query that selects either a single multi-dimensional relevant area (conjunctive query) or multiple ones (disjunctive query).

AIDE incorporates three exploration phases. First, we focus on collecting samples from yet unexplored areas and identifying single relevant objects (\emph{Relevant Object Discovery}). Next,  we strive to leverage single relevant objects to generate a user model that identifies relevant \emph{areas} (\emph{Misclassified Exploitation}). Finally, given a set of discovered relevant areas, we  gradually refine their boundaries (\emph{Boundary Exploitation}). In each iteration $i$, these three phases define the new sample set we will present to the user. Specifically, if $T^i_{d}$, $T^i_{m}$ and $T^i_{b}$  samples will be selected by the object discovery, the misclassified and the boundary exploitation phase, then the user is presented with $S_i = T^i_{d} + T^i_{m} + T^i_{b}$ samples. 

Our three exploration  phases  are designed to collectively increase the accuracy of the exploration results. Given a set of relevant objects from the object discovery step, the misclassified exploitation  increases the number of relevant samples in our training set while reducing the misclassified objects (specifically false negatives). Hence, this step improves both the recall and the precision parameters of the $F$-measure metric.  The boundary exploitation further refines the characterization of the already discovered relevant areas. Therefore, it discovers more relevant objects and eliminates misclassified ones, leading also to higher recall and precision.  Next, we discuss in detail each phase. More details about these exploration techniques can be found in~\cite{aide_sigmod14}.

%

\subsection{Relevant Object Discovery}\label{s:exploration}

Our first exploration phase aims to discover relevant objects by showing  to the user samples from diverse data areas. 
To maximize the coverage of the exploration space we follow  a well-structured approach that allows us  to 
 (1) ensure that the exploration space is explored widely, (2) keep track of the already explored sub-areas, and (3) explore different data areas in different granularity.

Our  approach  operates on a set of {\em hierarchical exploration grids}. Given an exploration task on $d$ attributes, we define the  \emph{exploration space} to be the $d$-dimensional data area defined by the \emph{domain} of these attributes.  AIDE creates off-line a set of grids and each grid divides the exploration space into $d$-dimensional  cells with  equal width in each dimension.  We refer to each grid as an  \emph{exploration level} and each level has a different  granularity, i.e., cells of different width. The lower the exploration level the more  fine-grained the grid cells (i.e., smaller cells) it includes and therefore moving between levels allows us to ``zoom in/out'' into specific areas as needed.

\begin{figure}[t]
\centering 
 \includegraphics[totalheight=1.7in, angle=-90] {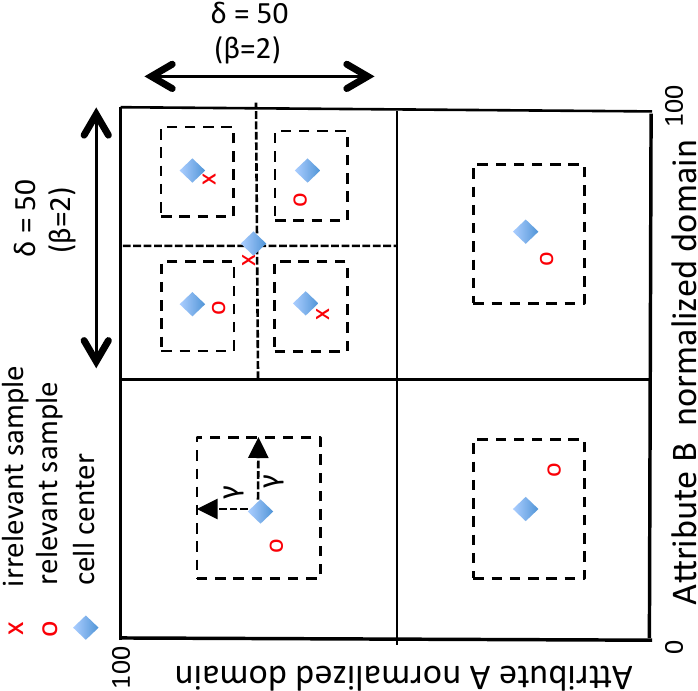}
\caption{{\small Grid-based object discovery example in 2-D space.}} 
\label{f:grid}
\vspace{-5mm}
\end{figure}

{\bf Exploration Level Construction} To generate an exploration level on a $d$-dimensional exploration space we  divide each normalized attribute domain\footnote{We normalize each domain to be between [0,100]. This allow us to reason about the distance between values uniformly across domains. Operating on actual domains will not affect the design of our framework or our results.} into $\beta$ equal width ranges, effectively creating $\beta^d$ grid cells. The $\beta$ parameter defines the granularity of the specific exploration level. A higher number leads to more grid cells of smaller width per dimension and  the use of more samples to explore \emph{all} grid cells for fine-grained search for relevant objects.
Each cell in our grid covers a certain range of attribute values for each of the $d$ exploration attributes. Therefore, each cell includes a set of unique attribute value combinations. 
Each  combination can be mapped to a set of data objects that match these attribute values.  Figure~\ref{f:grid} shows a two-level  2-dimensional hierarchical grid (we show the second  level only for the top right grid cell).  \cut{Next we present the general algorithm for this phase which explores uniformly all cells,  collects one  object from each cell and shows it to the user for feedback.}

{\bf  Discovery Phase}  Our  exploration  retrieves one data object from each non-empty cell. The goal is to uniformly spread the samples we collect across the grid cells to ensure the highest coverage of the exploration space. We achieve that by retrieving objects that are on or close to the center of each cell. Since the exploration levels are defined on the normalized domains of $d$ attributes and we split each domain to $\beta$ equal width ranges, each cell covers a range of $\delta=100/\beta$ of the domain for each attribute. For each cell, we identify the ``virtual'' center  and we retrieve a single random object within distance $\gamma<\delta/2$ along each dimension from this center (see Figure~\ref{f:grid}). This approach guarantees that at each exploration level the retrieved samples are within $\delta\pm (2\times \gamma)$ normalized distance from each other. 

The sampling distance around the center of each cell, i.e., the $\gamma$ parameter, is adjusted based on the density of the cell.  Sparse cells use a higher $\gamma$ value than dense ones to increase their sampling areas and hence improve the probability of retrieving a sample from the grid cell. This also reduces ineffective zoom in operations into the next  level for that cell which may happen if no object is retrieved in the current exploration level. 

 In the each iteration we focus on a specific exploration level (starting frm the highest one) and we show to the user one object for each cell. If no relevant object  is retrieved from one cell, we can safely infer that the \emph{whole} grid cell is not included in any relevant area. However, sub-areas of the grid could partially overlap with some relevant areas. Therefore, in the following iteration we ``zoom-in'' by repeating the above process in  a lower exploration level. Figure~\ref{f:grid} shows a scenario where AIDE discovered one relevant object in all but the top right  cell as well as the zoom in operation for this cell. Here, our sampling areas are the smaller four sub-cells inside the higher level cell.

\subsection{Misclassified  Samples Exploitation}\label{s:misclassified}

While the object discovery phase bootstraps the discovery of relevant objects, it extracts at \emph{most one} object of interest in each sampling area explored. In order to offer acceptable accuracy, decision tree classifiers require a higher number of samples from the relevant class.  AIDE employs the \emph{misclassified samples exploitation} phase which improves the accuracy our predictions  by increasing the number of  relevant objects in our training set. 

\begin{figure}[t]
\centering 
 \includegraphics[totalheight=2.7in, angle=90] {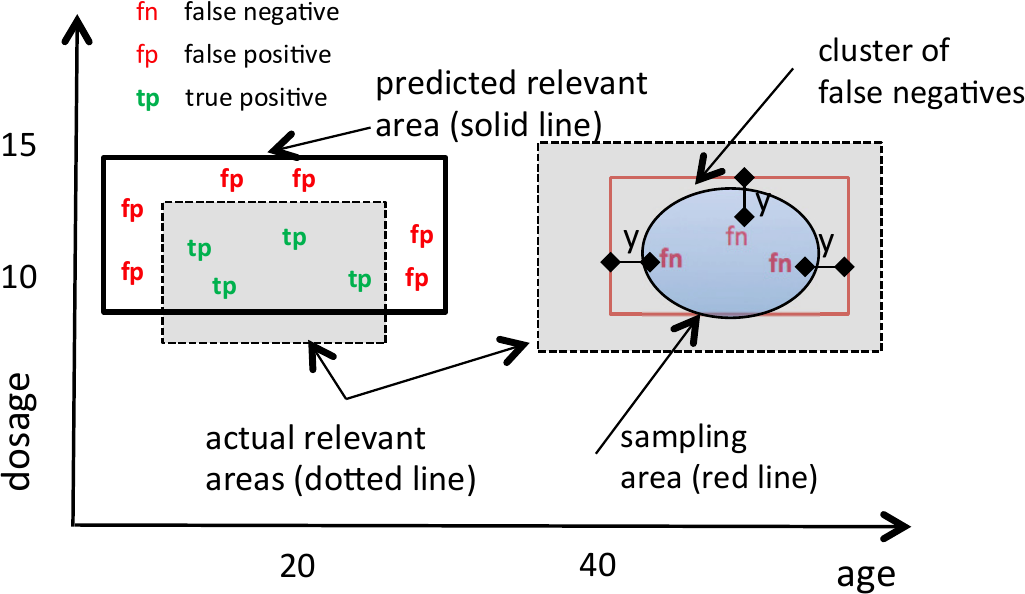}
 \vspace{-2mm}
\caption{{\small Misclassified objects and  cluster-based sampling.}} 
\label{f:misclassified_combo}
\vspace{-5mm}
\end{figure}

Misclassified objects can be categorized to: (i) \emph{false positives}, i.e., objects  that are categorized as relevant by the classifier but  labeled as irrelevant by the user and (ii) \emph{false negatives}, i.e., objects labeled as relevant but  categorized  as irrelevant by the classifier. False positives are less common because the classifications rules of decision trees aim to maximize  the homogeneity of their predicted  relevant and irrelevant areas~\cite{breinman84}. Practically, this implies that the classifier defines the relevant areas such as the relevant samples they include are maximized while minimizing the irrelevant ones. In fact, most false positives  are due to wrongly predicted boundaries of these areas. Figure~\ref{f:misclassified_combo} shows examples of false positives around  a predicted relevant area. Elimination of these misclassified samples  will be addressed by the boundary exploitation phase (Section~\ref{s:boundaries}).

False negatives on the other hand are  objects of interest that belong in an \emph{undiscovered}  relevant area. Examples of false negative are also shown in Figure~\ref{f:misclassified_combo}.  Relevant areas are  undiscovered by the decision tree due to insufficient  samples from within that area.  Hence, AIDE increases the set of relevant samples by collecting more objects  around  false negatives. 

{\bf Clustering-based Exploitation}
Our misclassified exploitation phase operates under the assumption that relevant tuples will be clustered close to each other, i.e., they typically form relevant areas. This implies that sampling around false negatives will increase the number of relevant samples. Furthermore, false negatives that belong in the same relevant area will be located close to each other. Hence, AIDE  generates  \emph{clusters of misclassified objects} and defines a new sampling area around each cluster. Specifically, it creates clusters using the \emph{$k$-means}  algorithm~\cite{breinman84} and defines one sampling area per cluster. An example of a  cluster of false negatives is shown in Figure~\ref{f:misclassified_combo}.

The main challenge in this approach is identifying the number of clusters we need to create. Ideally, we would like this number  to match the number of relevant areas we have ``hit'' so far, i.e., the number of relevant areas from within which we have collected at least one object. \cut{More cluster will lead us to oversample areas we have already ``hit'' while less clusters will lead us to under-sample again these areas.} We argue that the number of relevant objects created by the object discovery phase is a strong indicator of the number of relevant areas we have already ``hit''.  The object discovery phase  identifies objects of interest that belong to  different areas or the same relevant area.  In the former case, our indicator offers correct information. In the latter case, our indicator will lead us to create more clusters than the already ``hit'' relevant areas. However, since these clusters belong in the same relevant area they are typically close to each other and therefore the decision tree classifier eventually ``merges'' them and converges to an accurate number of relevant areas.

In each iteration $i$, the algorithm sets $k$ to be the overall number of relevant objects discovered in the object discovery phase.  Since our goal is to reduce the number of sampling areas (and therefore the number of sample extraction queries), we run the clustering-based exploitation only if  $k$ is less than the number of false negatives. Otherwise we collect $f$ random samples around \emph{each} false negative.  We collect  samples within a distance $y$ from the farthest cluster member  in each dimension. For each cluster we issue a query that retrieves $f \times c$ random samples within this sampling area, where $c$ is the size of the cluster (number of cluster members).   Our experimental results showed that $f$ should be set to a small number (10-25 samples) since higher values will increase the user effort without improving the exploration outcome. The closer the value $y$ is to the width of the relevant area we aim to predict, the higher the probability to collect relevant objects than irrelevant ones. \cut{An interesting optimization would be to dynamically adapt this value based on the current prediction of the relevant areas. We plan to explore this direction in our future work.  
}

\subsection{Boundary Exploitation}\label{s:boundaries}

Given a set of relevant areas identified by the decision tree classifier, our next phase aims to refine these areas by incrementally adjusting their boundaries. This leads to better characterization of the user's interests, i.e., higher accuracy of our final  results.  In this section, we describe our general approach. \cut{We also discuss a series of optimizations that 
allow us to reduce the number of samples and the sample extraction time required to refine the boundaries of already discovered areas.}

\begin{figure}[t]	
\centering 
 \includegraphics[totalheight=2.25in, angle=90] {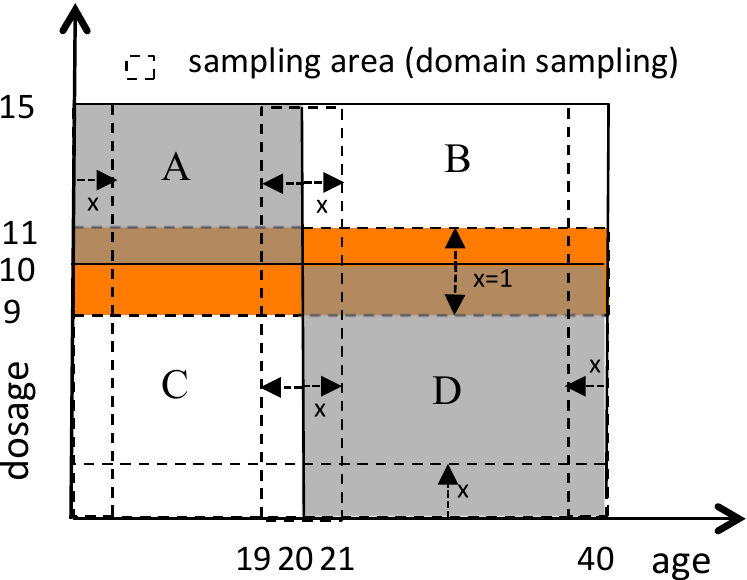}
 \vspace{-2mm}
\caption{{\small Boundary exploration for the relevant areas A and D.}} \label{f:boundaries_general}
\vspace{-5mm}
\end{figure}

 AIDE represents the decision tree classifier $C_i$ generated at the $i$-th iteration as a set of hyper-rectangles in a $d$-dimensional space defined by the predicates in $P_i^r \bigcup P_i^{nr}$, where the predicates ${P_i^r}$ characterize the relevant areas and predicates ${P_i^{nr}}$ describe the irrelevant areas. We iteratively refine these predicates by  \emph{shrinking} and/or \emph{expanding} the boundaries of the hyper-rectangles.  
 Figure~\ref{f:boundaries_general} shows the rectangles for the classifier in Figure~\ref{f:dtree}. If our classification is based on $d$ attributes ($d=2$ in our example) then a $d$-dimensional area  defined by  $p \in {P_i^r}$ will include objects  classified as relevant (e.g.,  areas A and D in Figure~\ref{f:boundaries_general}). Similarly, objects in an area defined by  $p \in {P_i^{nr}}$ are classified as irrelevant (e.g., areas B and C in Figure~\ref{f:boundaries_general}).

AIDE eliminates irrelevant attributes from the decision tree classifier by \emph{domain sampling} around the boundaries. Specifically, while we shrink/expand one dimension of a relevant area we collect random samples over the \emph{whole} domain of the remaining dimensions. Figure~\ref{f:boundaries_general} demonstrates our technique: while the samples we collect are within the range $11 \le dosage \le 9$ they are randomly distributed on the domain of the $age$ dimension.  \cut{Our experimental results are based on this approach and we observed that the quality of our final classifier was noticeably improved compared with an approach that selects samples bounded in all dimensions of the relevant areas.
 }

Our evaluation showed that this phase has the smallest impact on the effectiveness of our model: not discovering a  relevant area can reduce our accuracy more than a partially discovered relevant area with imprecise boundaries. Hence, we constrain the number of samples used during this phase to $\alpha_{max}$. This allows us to better utilize the user effort as he will provide feedback mostly on samples generated from the previous two, more effective phases.

\cut{Our approach aims to distribute an equal amount of user effort to refine each boundary. } Let us assume the decision tree has revealed $k$ $d$-dimensional relevant areas. Each area has $2^d$ boundaries. Hence we collect $\alpha_{max}/(k \times 2^d)$ random samples within a distance $\pm x$ from  each boundary.  This approach is applied across all the boundaries of  the relevant hyper-rectangles, allowing us to shrink/expand each dimension of the relevant areas. The new collected samples, once labeled by the user, will increase the recall metric: they will discover more relevant tuples (if they exist) and eventually refine the boundaries of the relevant areas.

The $x$ parameter can affect how fast we converge to the real relevant boundary. If the difference between the predicted and  real boundaries  is less than $x$,  this phase  will retrieve both relevant and irrelevant samples around the boundary and allow the decision tree  to more accurately predict the real boundary of the relevant area. Otherwise, we will mostly collect relevant samples. This will still improve our prediction of the boundary by bringing it closer to the actual one, but it will  slow down the convergence to the actual relevant area. We follow a conservative approach and set $x$ to  1 (we search for objects with normalized distance $\pm 1$ from the current predicted boundary). This gradually improves our recall. 

This phase includes a number of further optimizations, such as detecting and avoiding sampling overlapping areas as well as adjusting the number of samples to the convergence rate of the user model. These optimizations improve  AIDE's effectiveness and efficiency and they are described in detail in~\cite{aide_sigmod14}.

\section{Performance Optimizations}\label{s:optimizations}

In this section we describe a set of novel optimizations  we introduced in AIDE. These include techniques that: (a) handle exploration on skewed data distributions, (b) leverage the informativeness of  samples to improve AIDE's effectiveness,  (c) extend the expressiveness  of the user feedback model to accelerate the convergence to an accurate model and (d) reduce the size of our exploration space to offer highly interactive times. We note that the first three techniques are new optimizations that we added to the original version of AIDE introduced in \cite{aide_sigmod14,aide_vldb15}.

\subsection{Skew-aware Exploration}\label{s:explore_skewed}

Skewed data distributions are prevalent in virtually every domain of science. For example, in astronomy stars
and other objects are not uniformly distributed in the sky. Hence, telescope measurements have corresponding areas of density and sparsity. In our framework, skewed data distributions could hinder the discovery of relevant objects. This is due to the fact that our initial exploration step (Section~\ref{s:exploration}) is designed to distribute the number of collected samples evenly across this space.  While for uniform data distributions this approach will be effective, in the presence of skew, it could slow convergence to an accurate user model, since dense areas will be  under-sampled compared with the sparse ones. 
To address  this challenge we introduce a new sampling  technique designed to operate effectively on  both uniform and skewed data distributions, i.e.,  predict with high accuracy and less effort relevant areas that appear in either dense or sparse sub-areas of the exploration space.  

Specifically, we modified our first exploration phase to combine the grid-based exploration approach with a clustering technique that allows us to identify dense areas and increase our sampling effort within them.  Our technique uses the $k$-means algorithm~\cite{breinman84} to partition the data space into $k$ clusters and each cluster is characterized by its centroid while  database objects are assigned to the cluster with the closest centroid. Thus, each cluster includes similar objects (where similarity is defined by a distance function) and each  centroid serves as a good representative of the cluster's objects. Similarly to our grid-based exploration, we create multiple exploration levels, where higher levels include fewer clusters than lower ones. 

In parallel, AIDE maintains its grid-based exploration levels, i.e., it  divides the exploration space to $k$ sampling areas (i.e., grid cells)  of the same size independently of the distribution of the exploration space (as described in Section~\ref{s:exploration}). For uniform distributions, the  cluster-based and the grid-based sampling areas overlap. In this case, using  any of the two types of sampling areas is sufficient to discover relevant areas.  However, in the presence of skewed exploration domains most of the clusters will be concentrated to dense areas leaving sparse areas under-sampled. Maintaining our grid-based sampling areas allows us to sample also sparse sub-areas and discover relevant areas of low density. 

We now describe the details of our exploration technique. AIDE starts its exploration using the cluster-based exploration levels and   collects samples around the centroid of each cluster. Specifically, we select one object per cluster within distance $\gamma < \delta$  along each dimension from the cluster's centroid, where $\delta$ is the radius of the cluster. We initialize our exploration on the highest exploration level (i.e., with the few clusters). If no interesting objects are discovered, we sample the next level where finer-grained clusters are available. 

Next, AIDE uses the grid-based approach to sample sparse sub-areas. Here, we calculate the density value $s$ for every grid cell, where skew is defined as $s = {u}/{p}$, $u$ is the number of unique tuples mapped n that cell and $p$ is the number of all possible unique tuples that would exist if the cell was ``full'', i.e., there were to be  a tuple for each (attribute,value) combination of cell. The higher the $s$ value the denser a cell is. In this step AIDE only samples non empty sparse grid cells for which $s\leq t$. To set the $t$ value we analyse offline the distribution characteristics of  our exploration space to identify sparse sub-areas and we set $t$ to  the average density of the grid cells covering those areas.  \cut{To set the parameter $t$ we apply the $k$-means clustering algorithm offline on different exploration spaces (i.e., attribute combinations). For each exploration space we calculate the average skew value of the cells that would only get explored in lower exploration levels if we were using the cluster-based exploration levels.} AIDE   samples these sparse cells by extracting one random sample close to the center of the grid cell. 

The user is presented by the samples collected by both the grid-based and the cluster-based sampling areas. This hybrid approach allows us to adjust our sample size to the skewness of our exploration space (i.e., we collect more samples from dense sub-areas) while it ensure that   any sparse relevant areas will not be missed (i.e., sparse sub-areas are sufficiently explored).

\subsection{Probabilistic Sampling}\label{s:uncertainty}

AIDE relies on an pool-based active learning paradigm for discovering user interests, i.e., samples are picked from a pool of unlabeled data objects and presented to the user for labeling. Existing pool-based sampling strategies~\cite{Settles10activelearning}  exhaustively examining \emph{all} unlabeled objects available, searching for the best  sample to show to the user.  Clearly such an approach cannot scale on big dataset. AIDE addresses this challenge by identifying a small number of sub-areas of the total exploration space to sample and within each area it collects  \emph{random} samples, therefore eliminating greedy sampling techniques.  
%
%

While random sampling is highly effective especially in the boundary exploitation step (e.g.,  it distributes the samples across the whole domain of our exploration attributes which eliminates irrelevant attributes from the classifier, see Section~\ref{s:boundaries}), it suffers from certain limitations. In particular, in the misclassified exploitation phase, random sampling treats each samples uniformly and it does not leverage the informativeness of the samples, which could  potentially lead faster to an accurate user model. In other words, random sampling does not answer the question ``which candidate samples to show to the user in order to reduce the total number of labeled samples needed for learning''. To address this question, AIDE includes a new \emph{probabilistic sampling} strategy for the misclassified exploitation phase.

\cut{Both in the misclassified exploitation phase and the boundary exploitation phase the samples that we select to present to the user are randomly chosen from certain strategic areas. In the misclassified exploitation phase we collect $f$ random samples from within a distance $y$ from the farthest cluster member in each dimension for each of the clusters that are generated during the phase. In the boundary exploitation phase we collect $\alpha_{max}$ random samples within a distance $\pm x$ from each boundary of each hyper-rectangle that we have predicted.}

Active learning has proposed a number of sample selection approaches that evaluate the informativeness of unlabeled samples~\cite{Settles10activelearning}. In all these strategies the informativeness of a sample (e.g., the probability of being relevant or not) is either generated from scratch or sampled from a given distribution. In our framework, we do not assume a known distribution for our relevant or irrelevant objects. Instead we leverage the user's relevance feedback to calculate for each unlabeled object its informativeness,  i.e., its probability of being labeled as relevant or irrelevant (aka  posterior probability). Given this probability, we use the \emph{uncertainty sampling} strategy to identify the next set of samples to show to the user.

We now discuss how evaluate the posterior probability of unlabeled samples, given a set of relevant samples $S^+$ and irrelevant samples $S^-$.  AIDE considers each labeled sample as basis for a nearest neighbour classifier with only one training sample and consider each unlabeled object to be a test example that has to be classified into the  relevant or non-relevant class. We then combining these classifiers in order to ``blend'' information from all the user's collected feedback~\cite{deselaers_08, combineclassifier_98}.

Formally, we assume that, given a sample labeled as relevant by the user $s_{+}$, the probability that a unlabeled sample $x$  is relevant ($r$) is: 
\[p_{x}(r|s_{+}) \propto \exp(-similarity(x,s_{+}))\]
where $similarity(x,s_{+})$ return the similarity value between $x$ to $s_{+}$. Intuitively, this formula indicates that the probability of a sample $x$ being relevant increases exponentially with its similarity  to the relevant sample $s_{+}$. This is in accordance to our former argument that relevant samples will be clustered together in the exploration space and will be forming relevant areas.

 Analogously, we assume that the probability for a
sample $x$ being non-relevant ($n$)  increases exponentially when the sample is similar to a sample $s_{-}$ labeled as non-relevant by the user: \[p_{x}(n|s_{-}) \propto \exp(-similarity(x,s_{-})).\]

To calculate the posterior probability of a sample $x$ being relevant, we combine the individual classifiers from the set of relevant samples $S^{+}$  and the sets of irrelevant samples $S_{-}$ by using the sum rule~\cite{combineclassifier_98}. Specifically, given that $p_{x}(r|s_{+}) = 1-p_{x}(n|s_{-})$, 
\cut{Thus to calculate the probability of sample $x$  being relevant we can combine into one function the posterior probability of this sample being relevant given a set of relevant samples $S^{+}$ and the posterior probability of the same sample being non-relevant given a set of non-relevant samples $S^{-}$ :}

\begin{eqnarray*}
p_{x}(r|(S^{+}, S^{-})) &= 
\frac{\alpha}{|S^{+}|} \sum_{s_{+}\in S^{+}} p_{x}(r|s_{+}) + \\& \frac{1- \alpha}{|S^{-}|} \sum_{s_{-}\in S^{-}} 1 - p_{x}(n|s_{-})
\end{eqnarray*}
where $\alpha$ is a weighting factor we added to allow us to change the impact of the relevant and non-relevant samples. In the above formula if $\alpha=1$ we only take into account its distance from the set of relevance samples to calculate its posterior probability. In the opposite case if $\alpha=0$ we only consider its distance from the set of samples that are labeled as  non-relevant.

Given the posterior probability of a sample, we use the \emph{uncertainty sampling} strategy to select which samples to show to the user~\cite{Settles10activelearning}. In uncertainty sampling the user is presented with samples for which the classifier is the most uncertain about. When using a binary classification model, like in our case, uncertainty sampling selects the sample whose posterior probability of being positive is nearest to 0.5~\cite{Settles10activelearning}. These are the samples that we are the least certain about their relevance.

We apply this approach in our misclassified exploitation phase as follows. Our sampling areas are defined around the clusters of misclassified we have identified. Specifically, given a cluster of size $c$  we retrieve all samples within a distance $y$ from the farthest cluster member in each dimension.
Next, we calculate the posterior probability for each of these samples and we present to the user  $f \times c$ samples whose probability is closest to 0.5, where $f$ is our estimation of the number of relevant areas not identified by the classifier (see Section~\ref{s:misclassified} on how this number is calculated). Employing this technique allows us to discover the user's relevant area with less labeled samples proving the hypothesis that some samples are more informative than others. 



\subsection{Similarity Feedback Model}\label{s:maybe}

In our previous paragraphs we introduced exploration techniques that rely on binary relevance feedback, i.e., the user indicates whether the sample is relevant or not to her exploration task.  However, there exist numerous scenarios where although the user  cannot decidedly classify the relevance of an object,  she can indicate whether this object is ``close'' to her interests.
This label can be used when the user finds relevant some characteristics of the object but not necessarily all of them or if she is still uncertain about the relevance of the object, which is often the case when the user is unfamiliar with the underlying data set. 

Let us consider the case of a scientist exploring an astronomical dataset searching for  clusters of sky objects with unusually high brightness. Initially,  the user will be able to label star objects with high brightness values as potentially interesting. However, her understanding of which brightness values are in fact unusual crystallizes only after she has examined numerous sky objects of various brightness values. After that point she can identify unusually bright sky objects and label them as relevant. In another example, medical professionals searching for clinical trials for diabetes type A on 2 year old children can indicate that studies on diabetes type B  on 3 year old children are also of possible interest to her (e.g., since the symptoms, medication and side effects for  2 and 3 year old children can be quite similar). However, she will label as relevant only trials on 2 year olds. 

In the technical level, using a binary feedback model imposes a number of limitations to AIDE.  In the previous example let's assume the user labels trials on 3 year old children as relevant (since it is close to the age of the actual patient).  This will lead to less accurate classification model (e.g., AIDE will steer the exploration to studies on 3 year olds).  While  the use can modify this label in subsequent iterations, this will slow the convergence of the exploration to an effective  classification model. On the other hand labeling these trials as irrelevant does not capture the similarity of  these trials to the actual relevant objects (e.g., studies on 3 years are closer in the exploration space to the relevant trials than studies on 10 year olds). This similarity information, if expressed,  could  lead AIDE to focus its exploration on small ages and converge to an accurate model with less user effort. 

Furthermore, the similarity feedback could help improve AIDE efficiency when  predicting small areas of interest. In our current  approach, the smaller the relevant area we aim to predict, the higher  user effort (i.e., number of labeled samples) is required. This is a practical challenge especially when the relevant objects are clustered within very small  areas in the exploration space.  The smaller the relevant area the more zoom-in operations AIDE will execute in order to discover a relevant sample from within that area (Section~\ref{s:exploration}). These operations result in sampling more areas (i.e., grid cells), which increases the user effort as well as the number of sampling queries processed. A more expressive  feedback model that allows users to indicate that a sample is ``close'' to a relevant object could help us direct our zoom-in operations to only promising sub-areas of the exploration space. This will lead to an accurate user model with less user effort and exploration overhead.


To address the above challenges, we extended our user feedback model as follows. Users can indicate that an object is ``close'' to her interests by annotating it as a ``similar'' sample. This label should be used for samples with at least one attribute value that appears interesting or similar (``close'') to an relevant value.  The user  has the option to indicate these attributes, i.e., the dimension on which she found the sample to be interesting (e.g., age range in the above medical example, brightness in the scientific example). The system can then utilize this extra information to expedite the exploration process.  We note that our ``similarity'' annotations do not constitute a new label for our classification model, i.e., our decision tree classifier will continue to generate classification rules that predict only the relevant and irrelevant classes. Next, we describe our technique.

\cut{Ideally, we would like the user to express interest in objects similar (located close) to her interesting ones so that we can discover her relevant area without moving to lower exploration levels. To allow the user to express interest in samples that are close to the relevant area we extended our relevance feedback model to include one more label besides relevant and non-relevant: a \emph{weak-relevant} label. The main idea is that when we present the user with a sample that is very similar to the samples in the relevant object set $T^r$ the user can now label it as weak-relevant. The system can then utilize this extra information to expedite the exploration process of discovering the relevant area that is hidden nearby.
}


\begin{figure}[t]	
\centering 
 \includegraphics[totalheight=1.5in, angle=0] {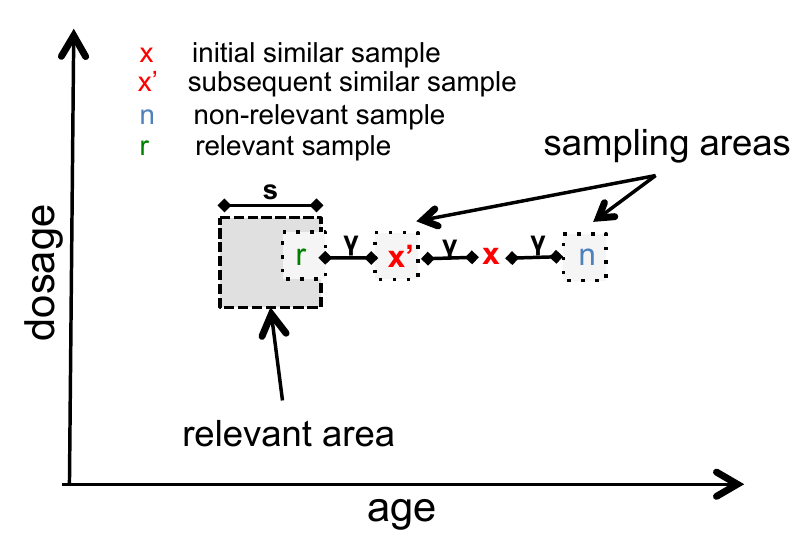}
\vspace{-2mm}
\caption{{\small {Similarity feedback exploration}.}} \label{f:weakRelevantFigure}
\vspace{-5mm}
\end{figure}

 \cut{To handle the case of encountering weak-relevant samples in our labeled sample set we deploy a new phase called \emph{weak-relevant sample exploitation}.}
{\bf Extended Feedback Exploration} We introduce one more exploration phase that defines sampling areas around each ``similar''  sample. Based on the definition of this label, each such sample $x$  is potentially ``close'' to a relevant object in at least one of the exploration dimensions. AIDE by default assumes that this similarity may be present in all dimensions unless the user explicitly indicates for which  dimensions she discovered similar values. We refer to these as the \emph{interesting dimensions}.  

Let us assume a sample $x$ annotated as ``similar' across a set of interesting dimensions $d$ (which are a subset of the set of exploration dimensions).  AIDE explores all possible interesting dimensions around  $x$ on the $d$ dimensional space aiming to identify relevant samples. Specifically,  there are $2^d$ possible exploration directions around the sample, i.e., for each dimension we explore both  higher and lower values of the $x$'s value on this dimension. Hence, we define $2^d$ sampling areas and we select one random sample close to the center of each  areas to present to the user. In Figure~\ref{f:weakRelevantFigure} we show a scenario of a 2-dimensional exploration space (age and dosage from our medical example), where the user has indicated as interesting only a single dimension (age). Hence, we have created 2 sampling areas around the sample $x$ and we have selected one sample within each of these  areas. 


We define the sampling areas to be located in a distance $\pm \gamma$ from the ``similar'' sample $x$ in each interesting dimension. If one of the new samples we present to the user is now closer to the relevant area we can expect  that the user will annotate it as a ``similar'' sample too. In the opposite case, we assume that the user will naturally be dissatisfied with these samples and will label them as non-relevant. In Figure~\ref{f:weakRelevantFigure}, let's assume that $x$ is a study on 5 years olds. If the patient's age is 3 then samples with lower age groups (e.g., sample $x'$)  will be also annotated as ``similar'' while samples with higher age groups (e.g., sample $n$)  will be irrelevant.  

In each iteration, AIDE will collect samples around each ``similar'' sample steering our exploration closer to the relevant area at each step. Eventually one of the sampling areas will overlap with the relevant area and the user will label the sample we extract from that area as relevant. Hence, sampling in a distance $\gamma$ from $x$  bring us closer or inside the relevant area.

The effectiveness of our $\gamma$ value  correlates with the range size of the relevant area  $s$ in each dimension (see Figure~\ref{f:weakRelevantFigure}). Let us assume $\gamma \leq s$ for some dimension. Then in the next iteration we will either we will either: a) sample inside the relevant area or b) our sampling areas will keep getting closer to the relevant area. The first case leads directly to the relevant area. In the second case  we guarantee that we will ``hit'' the relevant range in that dimension  in $d/\gamma$ iterations and hence after $d/\gamma -1$ ``similar'' sample annotations, where $d$ is the distance of the sample $x$ from the relevant range. In the opposite case where $\gamma>s$ we might move towards the relevant area but miss the area altogether; intuitively, our "step" is so large that we ``jump'' over the relevant range and never sample within it. In this case we expect the user to label the new samples we will present to her as non-relevant samples since our sampling areas are fending away from the relevant area instead of approaching it. AIDE detects this scenario and restarts this exploration phase from the original $x$ sample but a lower $\gamma$ value. Using this pattern, we keep adapting our $\gamma$ value until we ``hit'' a relevant sample. Finally if the user is willing to give us a \emph{hint} about the minimum size of the relevant areas $s$ we can set the value $\gamma$ of our ``step'' to be equal to $\delta$, which guarantees to sample within the relevant range in exactly $d/\gamma$ iterations. 

\subsection{Exploration Space Reduction}\label{s:sparereduction}

\cut{
The previous optimization forces our sample extraction queries of this phase to execute random sampling across the whole domain of each attribute. Such queries are particularly expensive, since they need to fully scan the whole domain of all attributes. Even when covering indexes are used to prevent access to disk, the whole index needs to be read for every query, increasing the sampling extraction overhead. We observed that although sampling on the whole domain improves our accuracy by an average of 42\%, it also has a time overhead of 95\% more than the other two phases. This overhead becomes even higher as we increase the size of our database. 
}

Our exploration techniques rely on sending a sampling query to the back end database system for each defined sampling area.
Such queries can particularly expensive. This especially true for the sampling queries generated by the boundary exploitation phase since they need to fully scan the whole domain of all attributes. Even when covering indexes are used to prevent access to disk, the whole index needs to be read for every query, increasing the sampling extraction overhead. 

An interesting artifact of our exploration techniques is that their effectiveness  does not depend on the frequency of each attribute value, or on the presence of all available tuples of our database. This is because each phase executes \emph{random} selections within data hyper-rectangles and hence these selections do not need to be deterministic. Hence, as long as the domain value distribution within these hyper-rectangles is roughly preserved, our techniques are still equally effective. This observation allows to apply our exploration on a sampled exploration space. Specifically, we  generate our sampled data sets using a simple random sampling approach that picks each tuple with the same probability~\cite{Olken94randomsampling}. We then execute our exploration on this smaller sampled space. Since this data space  maintains the same value distribution of the underlying attribute domains, our approach offers a similar level of accuracy but with significantly less time overhead.

\cut{We note that  this optimization can be used for both the misclassified and the boundary exploitation phases. This allows our sample extraction time to improve, with the most positive impact coming from the boundary exploitation phase. We observed that operating on a sampled dataset with size equal to 10\% of that of  the  original data set can improve our boundary exploitation time by up to 83.4\% and  the time for  the misclassified exploitation phase by up to 74.5\%. 
}


\section{Experimental Evaluation}\label{s:experiments}

Next, we present  experimental results from a micro-benchmark on the SDSS dataset~\cite{SloanSky} and from a user study. 

\subsection{Experimental Setup: SDSS Dataset}

\begin{figure*}[t]
\centering
\subfigure[\scriptsize{Accuracy for increasing area size (1 area)}.]{
\includegraphics[totalheight=1.03in, angle=0]{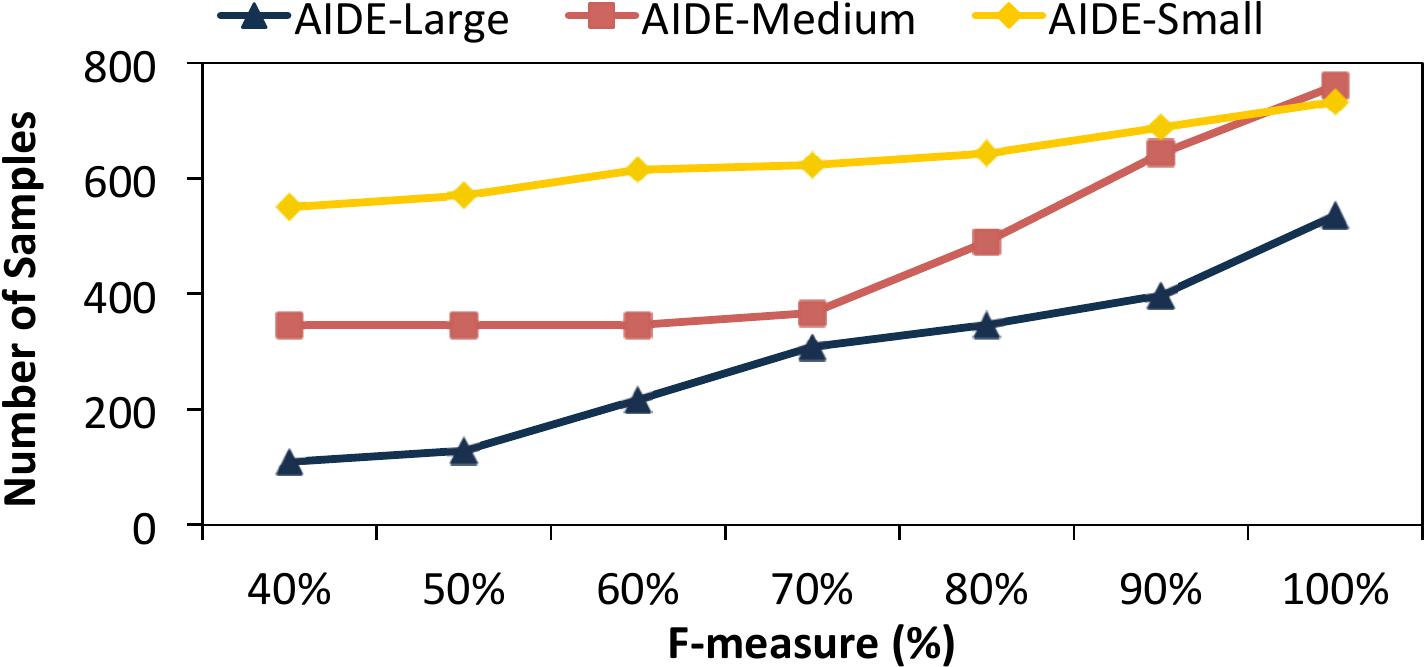}
\label{f:fmeasure_size}}
\subfigure[\scriptsize{Accuracy for increasing  number of areas (large areas)}.]{
\includegraphics[totalheight=1.03in, angle=0]{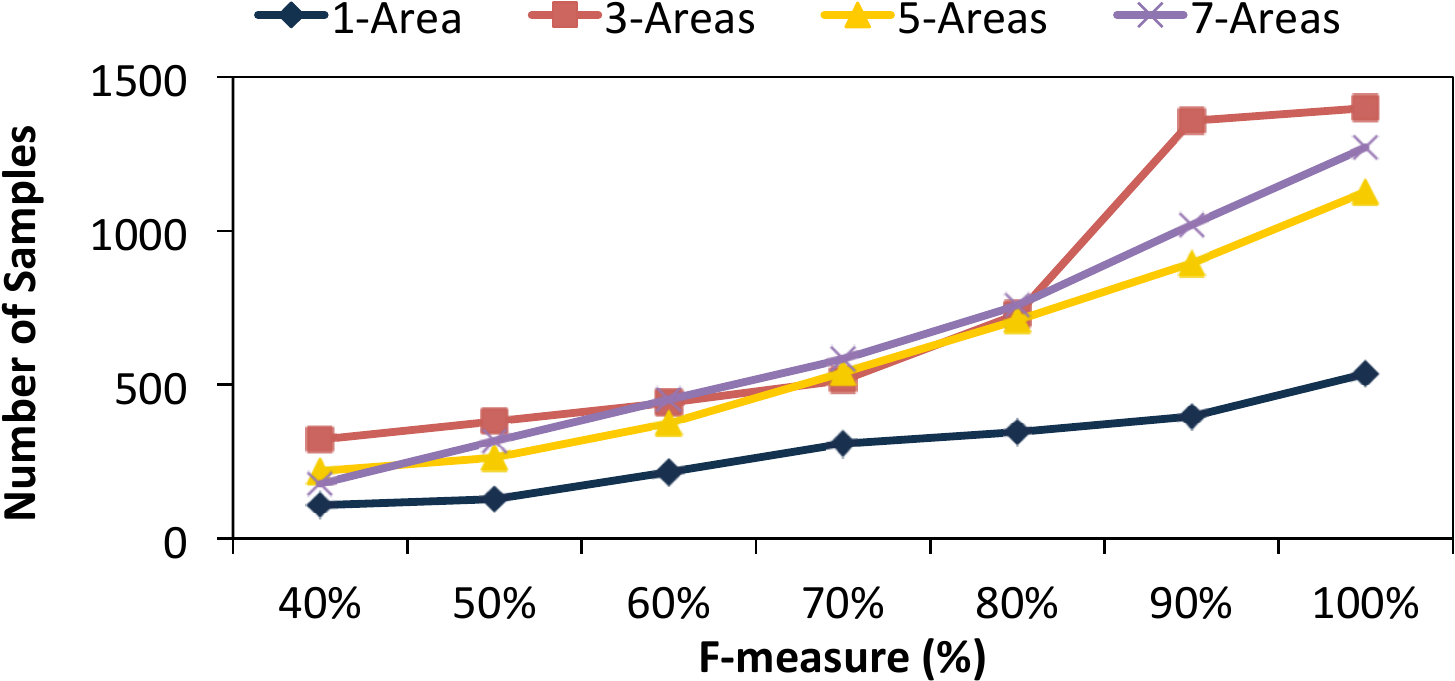}
\label{f:fmeasure_areas}}
\subfigure[\scriptsize{Time  for increasing  area size (1 area)}.]{
\includegraphics[totalheight=1.03in, angle=0]{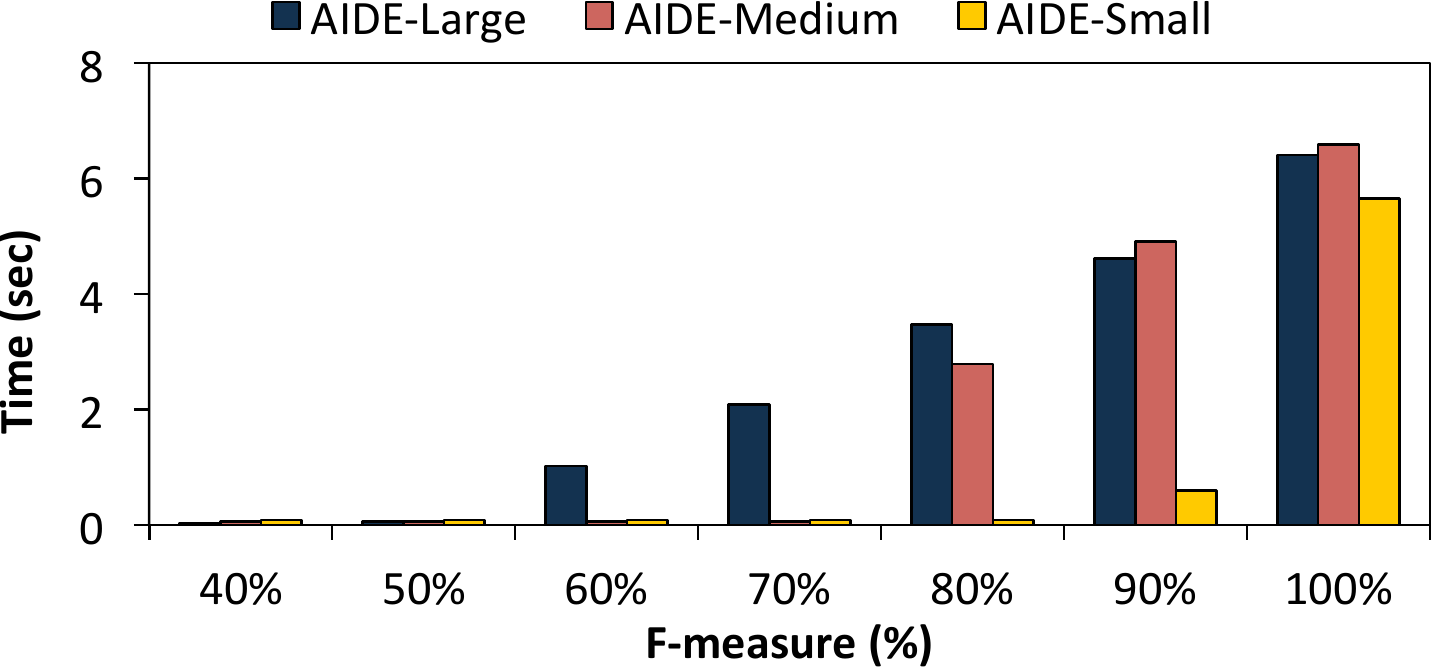}
\label{f:time_size}}
\subfigure[\scriptsize{Comparison to random exploration for increasing  area size ($>$70\% accuracy, 1 area).}]{
\includegraphics[totalheight=1.03in, angle=0]{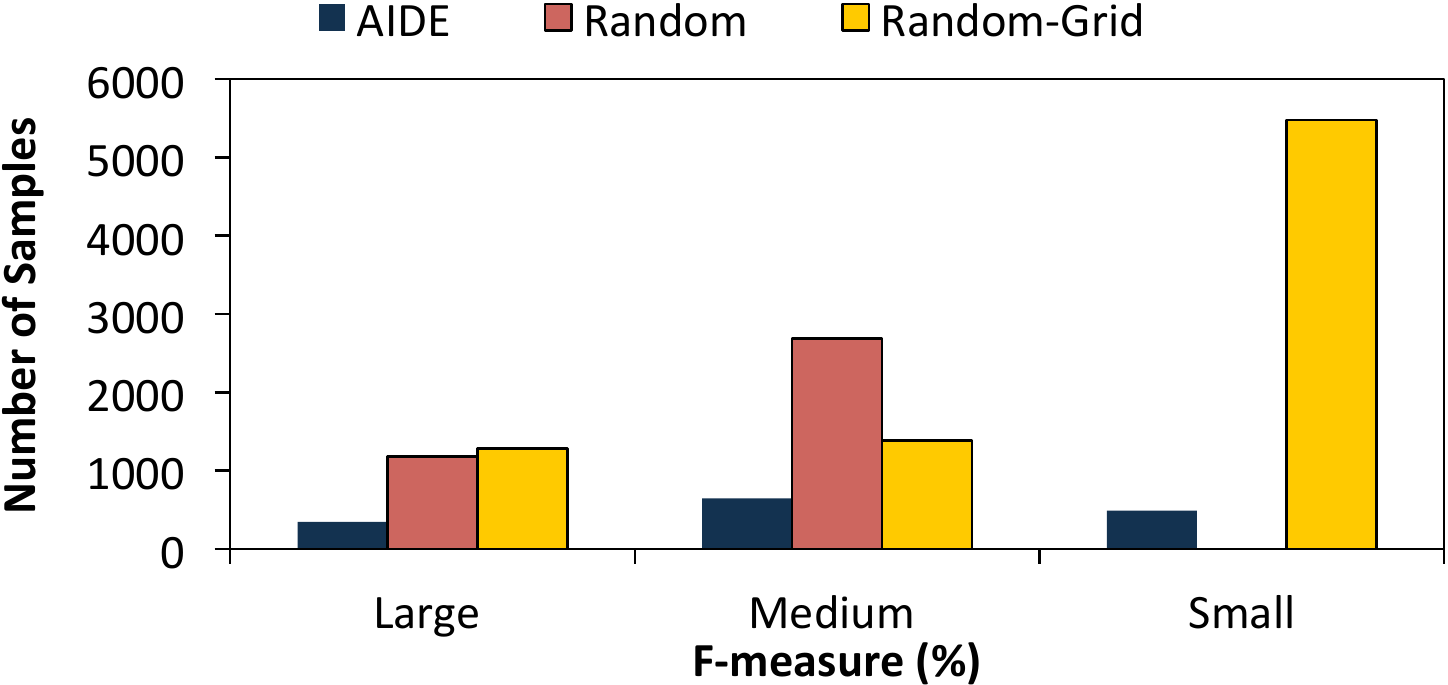}
\label{f:random_size}}
\subfigure[\scriptsize{Comparison to random exploration for increasing  number of areas ($>$70\% accuracy, large areas).}]{
\includegraphics[totalheight=1.03in, angle=0]{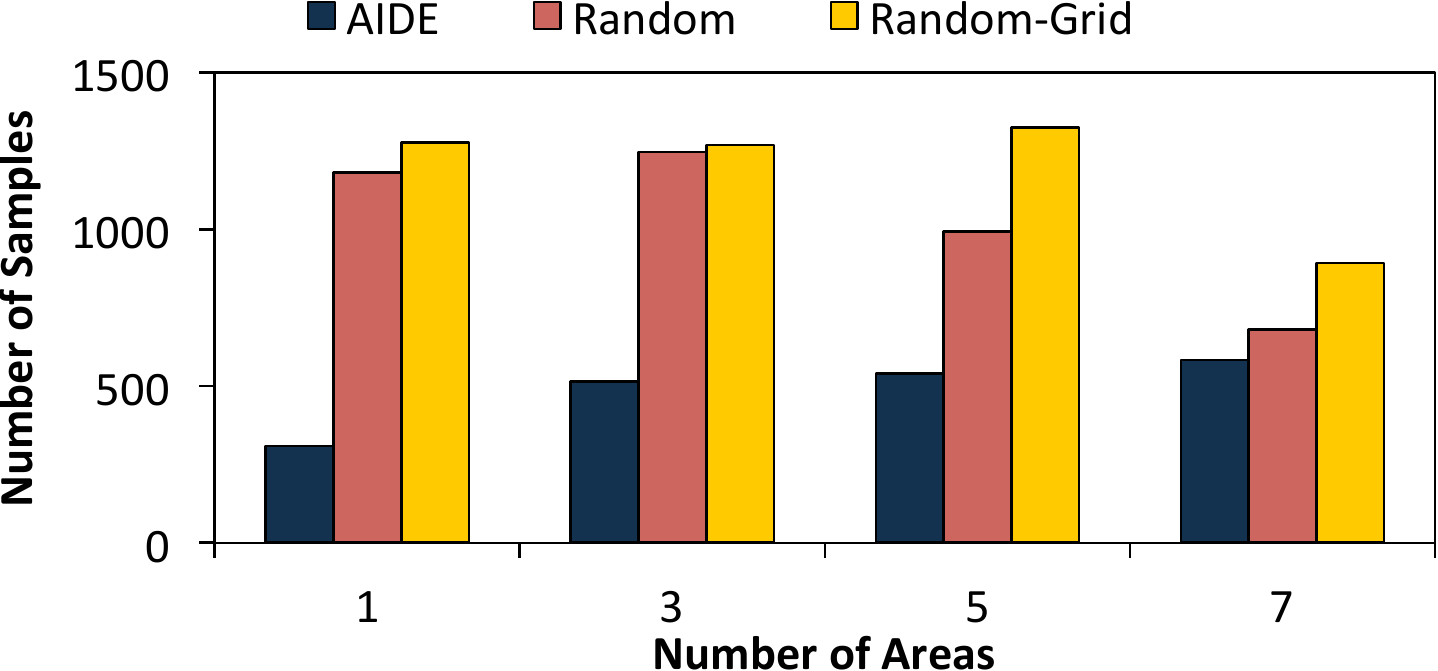}
\label{f:random_areas}}
\subfigure[\scriptsize{Impact of exploration phases (1 large area).}]{
\includegraphics[totalheight=1.03in, angle=0]{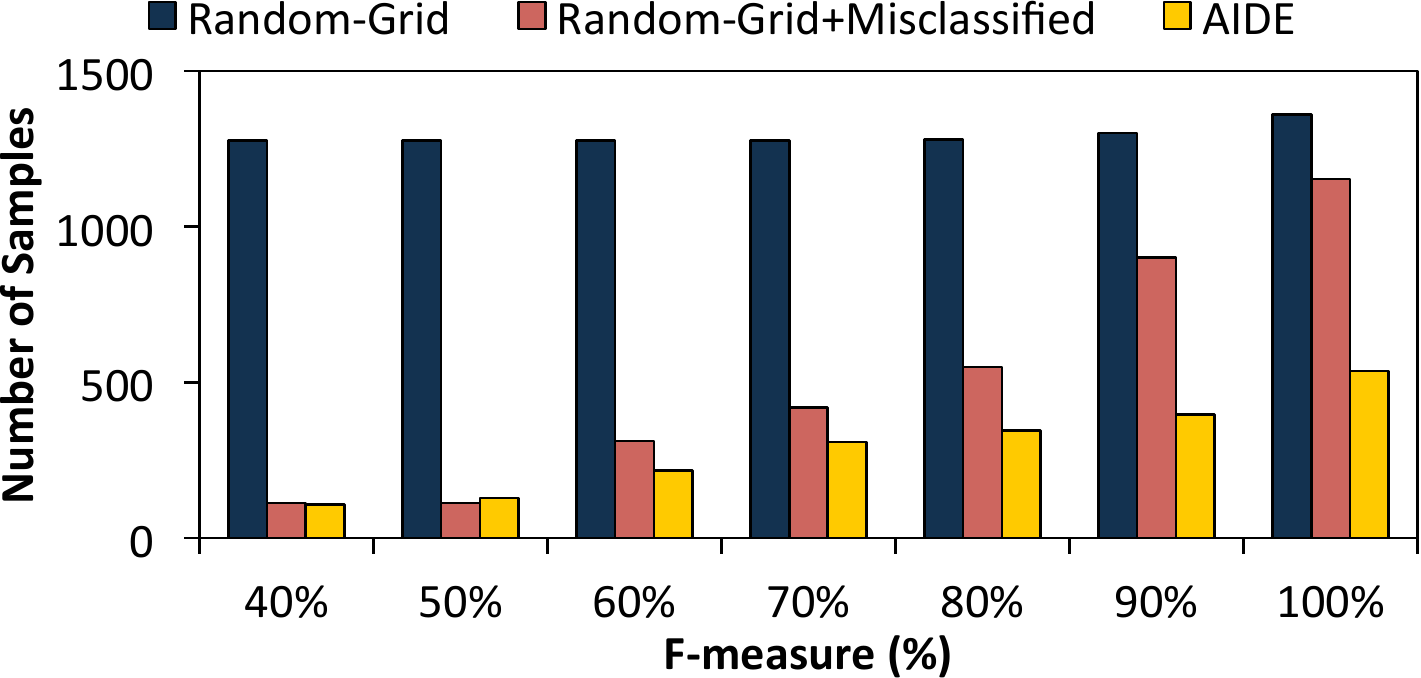}
\label{f:phase_impact}}
\vspace{-4mm}
\caption{\small{ Figures (a), (b) show AIDE's effectiveness, i.e., prediction accuracy. Figure (c) shows efficiency results, i.e., time overhead. Figures (d) and (e) compare  AIDE with random exploration techniques. Figure (f) demonstrates the effectiveness of AIDE's exploration phases.}}
\vspace{-1mm}
\label{f:effectiveness}
\end{figure*}

We implemented our framework on JVM 1.7. In our experiments we used various  Sloan Digital Sky Survey datasets (SDSS)~\cite{SloanSky} with a size of 10GB-100GB ($3\times 10^6 - 30 \times 10^6$ tuples). Our exploration was performed on  combinations of five  numerical attributes ({\tt rowc}, {\tt colc}, {\tt ra}, {\tt field}, {\tt fieldID}, {\tt dec})  of the {\tt PhotoObjAll} table. These are  attributes with different value distributions, allowing us to experiment with both skewed and roughly uniform  exploration spaces.  A covering index on these attributes was always used. We used by default a 10GB dataset and a dense exploration space on {\tt rowc} and {\tt colc}, unless otherwise noted. All our experiments were run on an Intel PowerEdge R320 server with 32GB RAM using MySQL.  We used  Weka~\cite{Weka_url}  for executing the CART~\cite{breinman84} decision tree algorithm and  the $k$-means clustering algorithm.  All experiments report averages of ten  exploration sessions. 

{\bf Target Queries} AIDE characterizes user interests and eventually ``predicts'' the selection predicates that retrieve her relevant objects. We focus on predicting range queries (we call them \emph{target queries}) and we vary their complexity based on: a) the  number of disjunctive predicates they include (\emph{number of relevant areas}) and b)  the data space coverage of the relevant areas, i.e., the  width of the range for each attribute (\emph{relevant area size}). Specifically,  we categorize relevant areas to \emph{small}, \emph{medium} and \emph{large}. Small areas have  attribute ranges with average width of 1-3\% of their normalized domain, while medium areas have width 4-6\% and large ones have 7-9\%. We also experimented with queries with a single relevant area (conjunctive queries) as well as complex disjunctive queries that select 3, 5 and 7 relevant areas. The higher the number of relevant areas and the smaller these areas, the more challenging is to predict them. 

The diversity of our target query set is driven by the query characteristics we observed in  the SDSS sample query set~\cite{SloanSky-Queries}. Specifically,  90\% of their queries select a single area, while  10\% select only 4  areas. Our experiments cover  even more complex cases of 5 and 7  areas.  Furthermore, 20\% of the predicates used in SDSS queries cover 1-3.5\% of their domain, 3\% of them have coverage around 13\%, and 50\% of the predicates have coverage 50\% or higher while the median coverage is 3.4\%. Our target queries have domain coverage (i.e., the relevant area size) between 1-9\% and our results demonstrate that we perform better as the size of the areas increases. Hence, we believe that our query set has a good coverage of queries used in real-world applications while they also cover  significantly more complex cases.

{\bf User Simulation} Given a target query, we simulate the user by executing the query to collect the exact \emph{target set }of relevant tuples. We rely on this set to label the new sample set we extract in each iteration as relevant or irrelevant depending on whether they are included in the target set. We also use this  set to evaluate the accuracy ($F$-measure) of our final predicted extraction queries.

{\bf Evaluation Metrics} We measure  the accuracy of our approach using the $F$-measure (Section~\ref{s:problemdef}) of our final data extraction predictions and report the  number of labeled samples required  to reach different accuracy levels. 
{Our efficiency metric is the \emph{system execution time} (equivalent to {\em user wait time}), which include the time for the space exploration, data classification, and sample extraction.
We may also report the total {\em exploration time}, which includes both the system execution time and  the sample reviewing time by the user. }

%

\subsection{Effectiveness \& Efficiency of AIDE}

Figure~\ref{f:fmeasure_size} shows AIDE's effectiveness when we increase the query  complexity by varying the size of relevant areas from large (\emph{AIDE-Large}) to medium (\emph{AIDE-Medium}) and small (\emph{AIDE-Small}). Our  queries have one  relevant area which is the  most common range query in SSDS. Naturally,  labeling more samples improves in all cases the accuracy. As the query complexity increases the user needs to provide more samples to get the same level of accuracy. By requesting feedback on only 215 out of  $ 3 \times 10^6$ objects AIDE  predicts large relevant areas with accuracy higher than 60\% (with 350 samples we have an accuracy higher than 80\%). In this case, the user needs to label only 0.4\% of the total relevant objects and 0.01\% of the irrelevant objects in the database.  Furthermore, AIDE needs  345 samples to predict  medium areas and  600 samples for small areas  to get an accuracy of at least 60\%.  
\emph{Hence, AIDE  decreases the user effort (i.e., reviewing objects) to a few 100's samples compared with the state-of-the-art ``manual'' exploration  which involves examining 1000's of objects (e.g., target queries return 26,817-99,671 relevant objects depending on the size of the relevant areas)}.

We also increased the query complexity by varying the number of areas from one (1) to  seven (7). Figure~\ref{f:fmeasure_areas} shows our results  for the case of large relevant areas. While AIDE can perform very well for common conjunctive queries (i.e., with one (1) relevant area), to accurately predict highly complex disjunctive queries more samples are needed. However, even for highly complex queries of seven (7) areas we get an accuracy of 60\% or higher with reasonable number of samples (at least 450 samples). 

Figure~\ref{f:time_size}  shows the execution time overhead (seconds in average per iteration). In all cases, high accuracy requires the extraction of  more samples which increases the exploration time. The complexity of the query (size of relevant areas) also affects the time overhead. Searching for larger relevant areas leads to more sample extraction queries around the boundaries of these relevant areas. However, our time overhead is acceptable: to get an accuracy of  60\%  the user wait time per iteration is  less than one second for small and medium areas, and 1.02 second for large areas, while to get highly accurate predictions (90\%-100\%) the user experiences 4.79 second wait time in average. To reach the highest accuracy ($>$ 90\%) AIDE executed 23.7 iterations in average for the large areas, 37 iterations for the medium and 33.4 iterations for the small areas. 

 {\bf Comparison with Random Exploration} Next  we compared AIDE  with two alternative exploration techniques. \emph{Random} randomly selects 20 samples per iteration, presents them to the user for feedback and then builds a classification model.  \emph{Random-Grid} is similar to Random but the sample selection is done on  our exploration grid, i.e., it selects one random sample around the center of each grid cell. This allows our samples to be evenly distributed across the exploration space. This approach also collects 20 samples per iteration. AIDE also limits  the  number of new samples  it extracts per iteration: we calculated the number of samples needed for the boundary and the misclassified exploitation and we used the remaining out of 20 samples to sample  grid cells. 

\begin{figure*}[t]
\subfigure[\scriptsize{{Accuracy for skewed data sets}  ( $>$70\% accuracy, 1 large area).}]{
\includegraphics[totalheight=1.15in, angle=0]{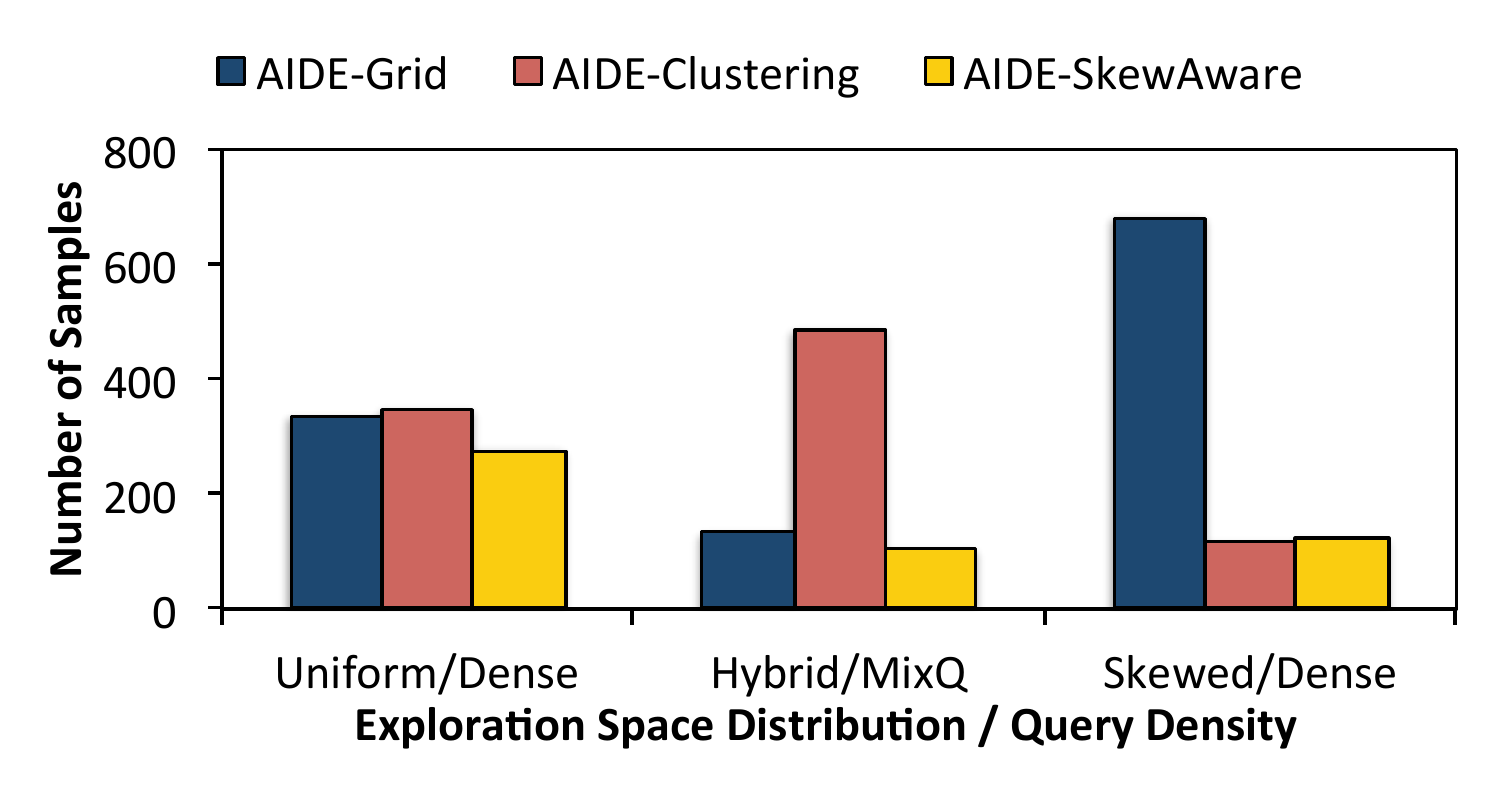}
\label{f:hybrid_fmeasure}}
\subfigure[\scriptsize{{Accuracy of probabilistic sampling} ($>$80\% accuracy, 1  area).}]{
\includegraphics[totalheight=1.15in, angle=0]{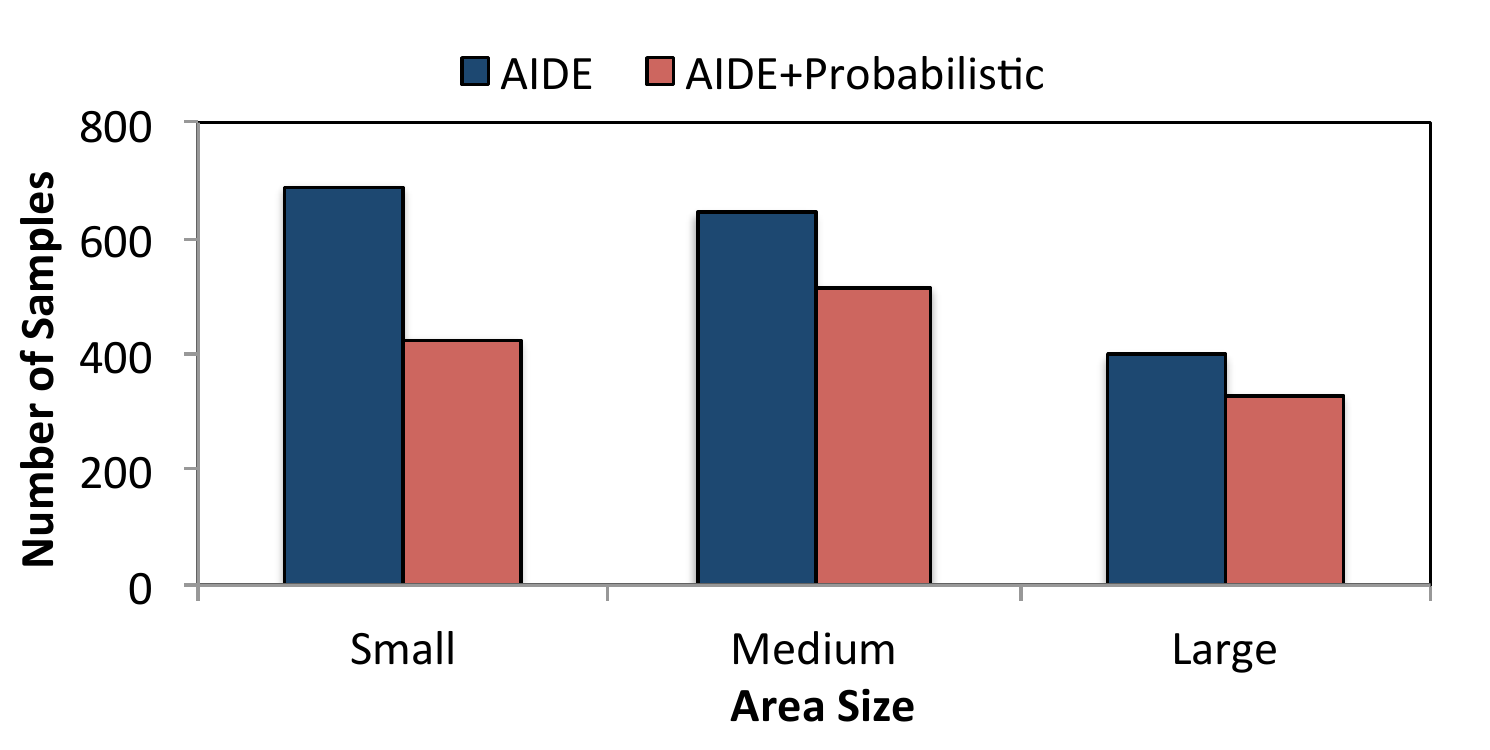}
\label{f:uncertainty_fmeasure}}
\subfigure[\scriptsize{{Time for probabilistic sampling ($>$80\% accuracy, 1  area).}}]{
\includegraphics[totalheight=1.15in, angle=0]{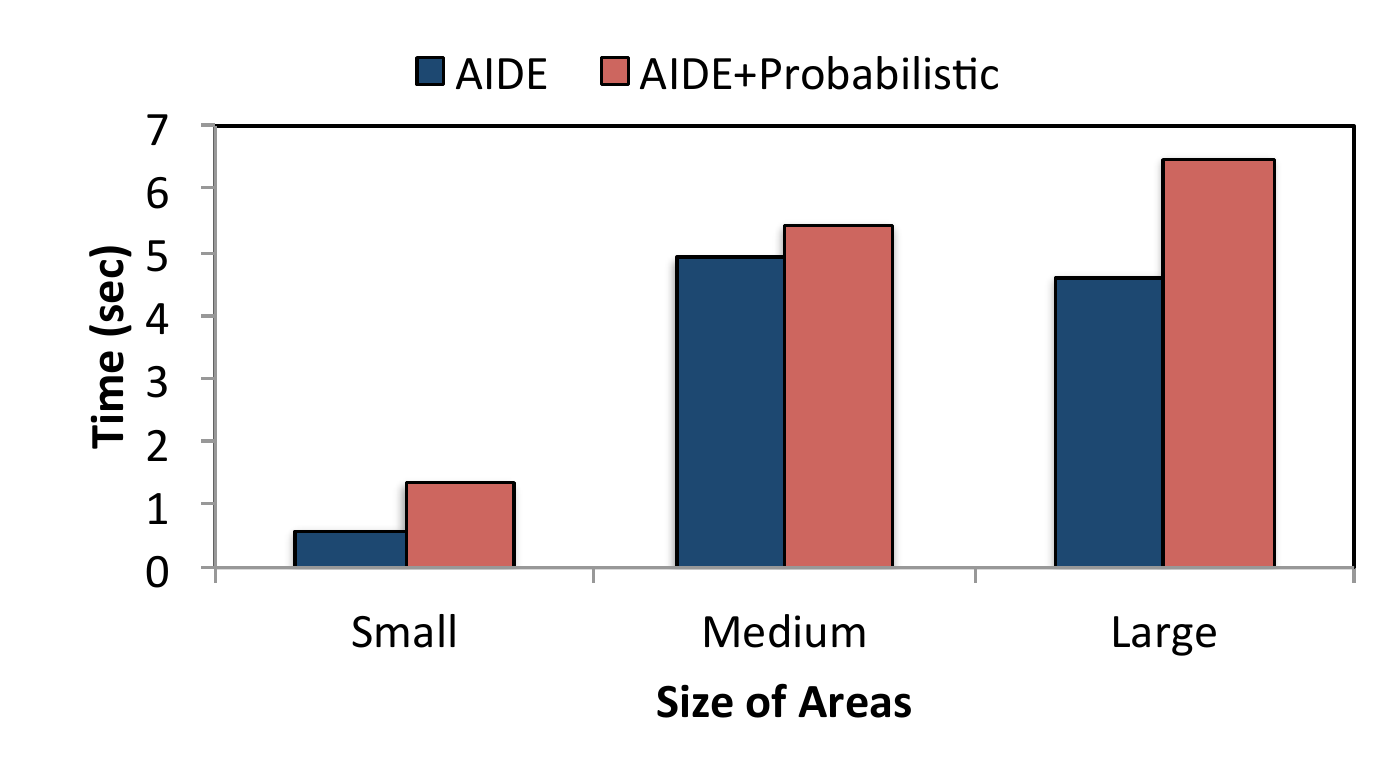}
\label{f:uncertainty_time}}
\subfigure[\scriptsize{Accuracy for extended  feedback model ($>$70\% accuracy, 1  area). }]{
\includegraphics[totalheight=1.15in, angle=0]{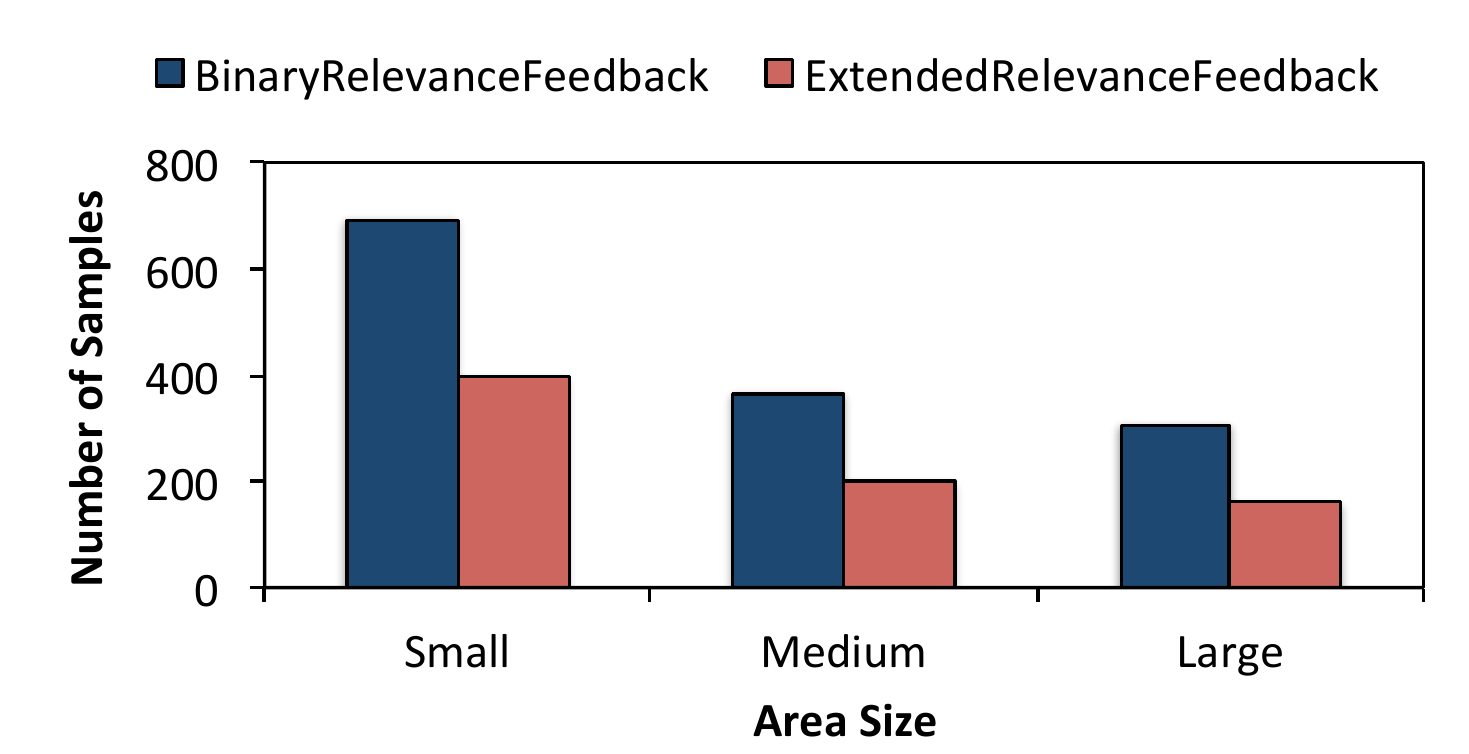}
\label{f:maybeFmeasure}}
\subfigure[\scriptsize{Accuracy for multi-dimensional exploration spaces  ($>$70\% accuracy, large areas).}]{
\includegraphics[totalheight=1.15in, angle=0]{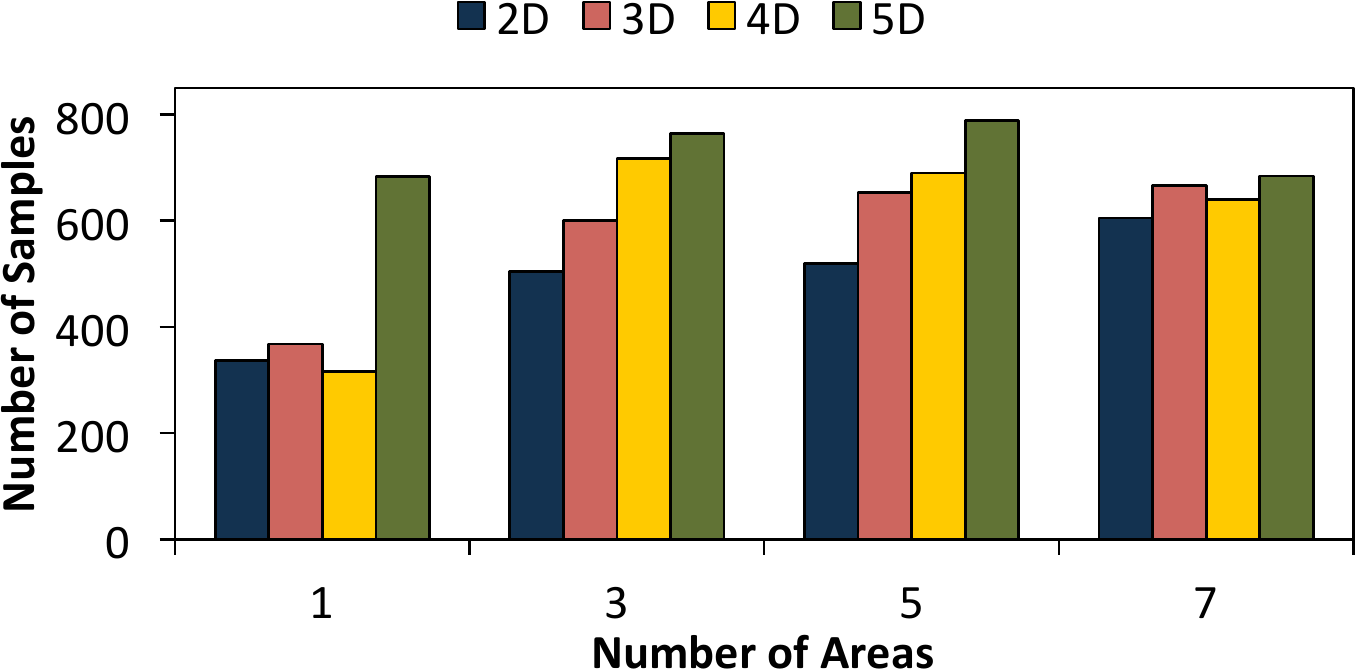}
\label{f:5D_fmeasure}}
\subfigure[\scriptsize{Time for multi-dimensional exploration spaces  ($>$70\% accuracy, large areas).}]{
\includegraphics[totalheight=1.15in, angle=0]{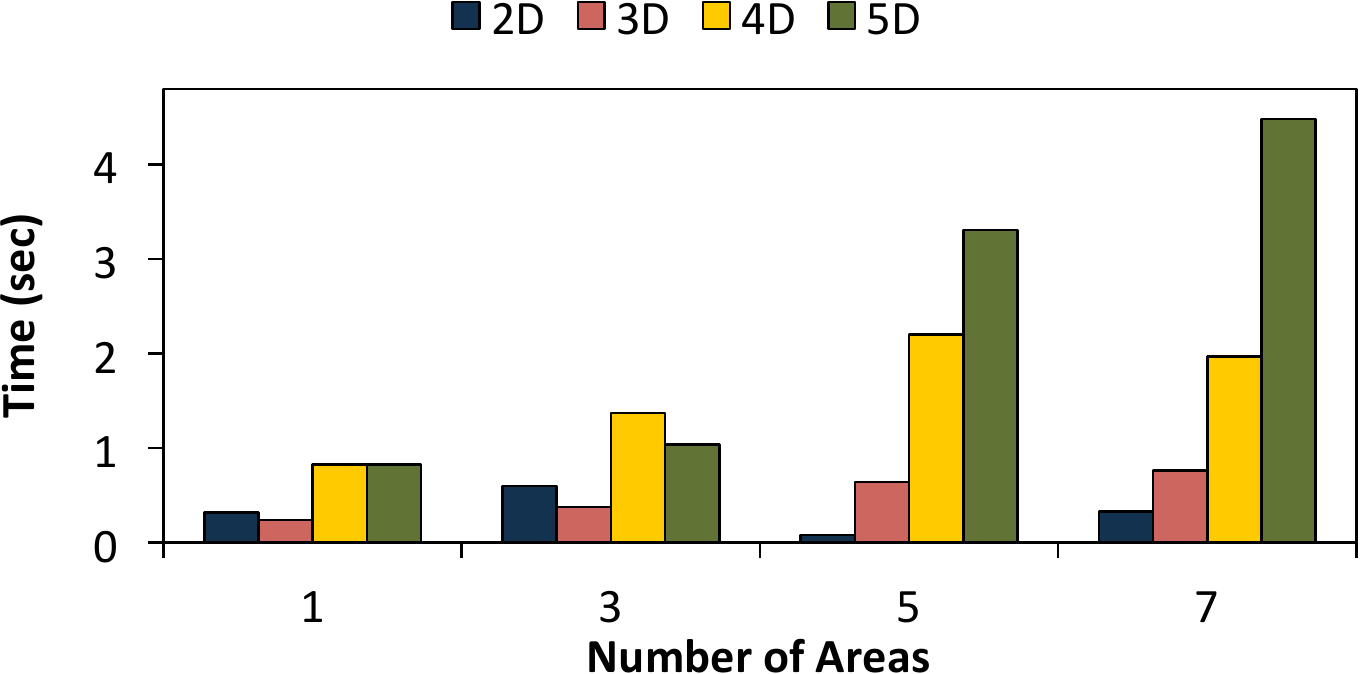}
\label{f:5D_time}}
\vspace{-4mm}
\caption{\small{Impact of performance optimizations: (a) Evaluation of the skew-aware exploration (Section~\ref{s:explore_skewed}),(b)-(c) Evaluation of the probabilistic sampling (Section~\ref{s:uncertainty}), (d) Evaluation the similarity feedback  model (Section~\ref{s:maybe}). Figures (e)-(f) study multi-dimensional exploration spaces.}} 
\label{f:optimizations}
\vspace{-4mm}
\end{figure*}


 Figure~\ref{f:random_size} shows the number of samples needed to achieve an accuracy of at least 70\% when our target queries have  one (1) relevant area and  varying  size. AIDE is consistently  highly effective: it requires only 308 samples for large areas and 365 and 623 samples in average for medium and small samples, respectively. Both random exploration approaches cannot discover small and medium areas with that few  samples. Random fails to discover  small areas of interest even when we increase the labeled set to 6,400 samples, while Random-Grid needs 5,457 samples in average for these complex queries. Random can identify medium and large  relevant areas with 70\% accuracy when given at least 2,690 and 1,180 samples respectively. Random-Grid is also highly ineffective, since it needs 1,380 and 1,275 samples in average for medium and large areas.  Figure~\ref{f:random_areas} shows the number of samples to achieve at least 70\% accuracy when varying the number of target relevant areas. AIDE consistently requires less samples (less than 500 samples for all cases) than Random and Random Grid (more than 1,000 samples in almost all cases).  \emph{Hence,  AIDE outperforms random sampling over all unlabeled objects since it samples only promising exploration sub-areas, leading to highly accurate results with less sampled data. }
 
 {\bf Impact of Exploration Phases} We also studied the impact of each exploration phase independently. Figure~\ref{f:phase_impact} compares the number of samples we need to reach different accuracy levels for queries with one large relevant area.  We compare  AIDE with two variants: one that uses  only the object discovery phase (\emph{Random-Grid}) and one that adds only the misclassified exploitation phase (\emph{Random-Grid+Misclassified}). The results  show that combining all three phases gives the best results.  Specifically,  using only the object discovery phase requires consistently more than 1,000 samples to get an accuracy greater than 40\%. Adding the misclassified exploitation phase reduces the sample requirements by 60\% in average while adding the boundary exploitation phase allows us to achieve higher accuracy with 42\% less samples in average.  \emph {Hence, combining all three phases is highly effective in  predicting relevant areas while  reducing the amount of user effort. }

\subsection{Skewed Exploration Spaces} 
 We also studied AIDE in the presence of skewed exploration spaces.  We experimented with three types of  2-dimensional exploration spaces: (a) \emph{Uniform} where we use two roughly uniform domains ({\tt rowc}, {\tt colc}), (b) \emph{Hybrid} that includes one skewed ({\tt dec}) and one uniform  domain ({\tt rowc}) and (c) \emph{Skewed} that uses two skewed domains ({\tt dec}, {\tt ra}).  We also experimented with the density of the target queries: (a) \emph{Dense}queries involve dense relevant areas  and (b) \emph{MixQ} queries cover both sparse and dense ranges of the relevant domains.  Figure~\ref{f:hybrid_fmeasure} shows the number of samples needed  to achieve accuracy greater than 70\% for  queries with one large  relevant area.  We  compare three variants of our system: (a) AIDE-Grid that uses the grid-based technique for the relevant object discovery phase,  (b) \emph{AIDE-Clustering} that uses only  clustering-based sampled for skewed distributions but not sampling within grid cells and (c) \emph{AIDE-SkewAware} that is a hybrid of the two previous techniques as described in Section~\ref{s:explore_skewed}. 

The results show that AIDE-SkewAware works  best under any combination of query density and exploration space distribution. When the distribution is uniform (Uniform) clusters and grid cells are highly aligned providing roughly the same results for all three techniques. Note that in this case all our relevant areas will be dense.  In the highly skewed data space (Skewed) we also used only dense relevant areas as the sparse areas were practically non populated. Here,  both the clustering-based technique and the skew-aware technique outperform the grid-based approach requiring 87\% less samples. This is because clusters are formulated in the dense sub-space while grid cells are created uniformly across the data space covering non populated exploration areas. This allows AIDE-Clustering and AIDE-SkewAware to sample smaller,  finer-grained areas than the grid-based approach, eliminating the need to zoom into the next exploration level.

Finally, for the case of hybrid distributions (Hybrid) we picked our relevant area to cover both dense ranges (for the uniform domain) and sparse ranges in the skewed domain, resulting to our mixed query case (MixQ). Here, the clustering technique creates most of its clusters on the dense areas and hence fails to discover relevant objects in the sparse ones. It therefore has to zoom into finer exploration levels  and it requires 73\% more samples  to converge to the same accuracy as the grid-based technique. However, AIDE-SkewAware samples both the dense areas where the clusters are located and the sparse areas which are covered by  the grid cell  and it discovers the relevant area. \cut{Furthermore, Figure~\ref{f:hybrid_time} compares the execution time overhead (seconds in average per iteration) for the three techniques. We can observe that all three techniques perform similarly in terms of efficiency. In all cases, the user wait time is fairly low; less than 2.1 seconds per iteration.}
 \emph{We conclude that combining sampling within clusters and grid cells is the best strategy for exploring both skewed  and non skewed domains}. 

\subsection{Probabilistic Sampling} 
Next, we examine the effectiveness and efficiency of the probabilistic sampling technique (Section~\ref{s:uncertainty}). In Figure~\ref{f:uncertainty_fmeasure} we measure the number of samples needed to reach an $F$-measure  greater than 80\% when the probabilistic sampling technique in the misclassified exploitation phase (\emph{AIDE+Probabilistic}). The experiments shows the results when we increase size of the relevant area from small areas to medium and large areas. \emph{AIDE requires less labeled samples to reach an accuracy when using the uncertainty sampling technique.} In average this new approach can reduce the user effort by 21\%. This confirms our hypothesis that some samples in the misclassified sampling area are more informative than others and they can be leveraged to improve the user's experience.

We also studied the overhead of this approach. In Figure~\ref{f:uncertainty_time} we can observe that the uncertainty sampling technique slightly increases our user wait time per iteration in all cases. This is because in each iteration we have to extract all samples within the sampling area, calculate its posterior probability and decide whether to present it to the user or not. The  user wait time per iteration was increased by 25\%  in average. However, in all cases the time overhead was less than 1.8 seconds which should not affect the user's interactive experience. This is because our technique searches for the most informative samples only within a small sub-set of the overall exploration space.

\subsection{Similarity Feedback Model} 
We also studied the effect of extending our relevance feedback model to include labels for similar but not necessarily relevant attributes.  Here, we label as ``similar'' samples that are within distance less than 10\% from an actual relevant object (this distance is measured in any of the exploration dimensions). Otherwise we label it as irrelevant.

 Figure~\ref{f:maybeFmeasure} compares AIDE's effectiveness when using  the binary feedback approach and the extended  feedback model. Here, we vary the size of the  target relevant area from small up to large and we measure the number of samples AIDE needs to reach an $F$-measure higher than 70\%.  \emph{The results indicate that annotating the similarity of objects can significantly reduce the labeling effort of the user}. This improvement is 42\% in average across all area sizes. This feedback is particularly useful  in the case of the  small relevant areas where the user effort can be significant. Here, the user's ``similar'' annotations  steer the exploration towards the direction of the relevant samples and the labeling effort is significantly reduced. We also measured the impact of our model on the user wait time and in all cases was under 0.1 seconds which should be unnoticeable by the user in our interactive system. We omit the graph due to space limitations. 

\cut{
Figure~\ref{f:maybeTime} compares the average user wait time per iteration when using  the binary and the extended relevance feedback model. As we can see, the improvement in user effort with the relevance feedback model comes with a small increase in the average user wait time per iteration. This overhead is due to the extra sampling queries that we need to execute during the extended feedback exploitation phase to sample the areas around our ``maybe'' samples. Specifically, the user wait time increased in average by 16\% but in all cases it was under 0.1 seconds which should be unnoticeable by the user in our interactive system. 
}

\subsection{Scalability}

\begin{figure*}[t]
\centering
\subfigure[\scriptsize{Accuracy for increasing data set size (1 large area).}]{
\includegraphics[totalheight=1.03in, angle=0]{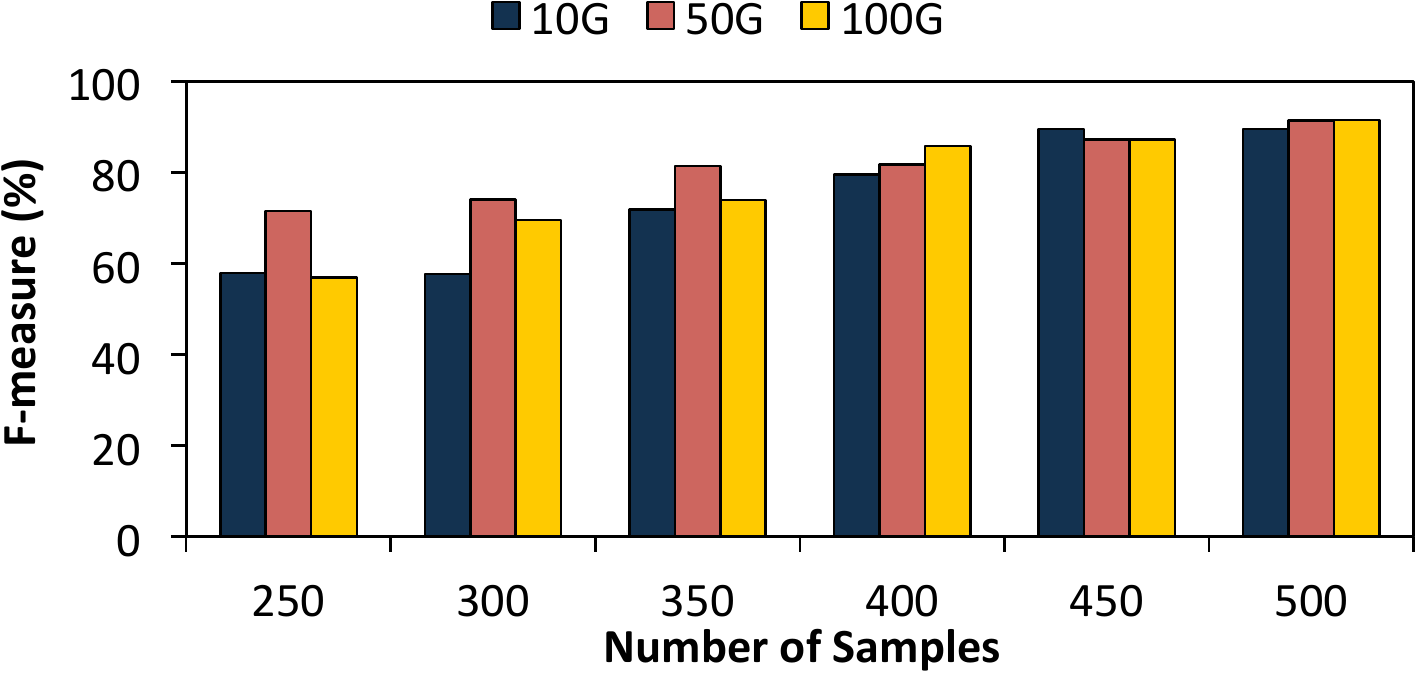}
\label{f:fmeasure_dbsize}}
\subfigure[\scriptsize{Impact of  space reduction on accuracy and time overhead (1 large area).}]{
\includegraphics[totalheight=1.03in, angle=0]{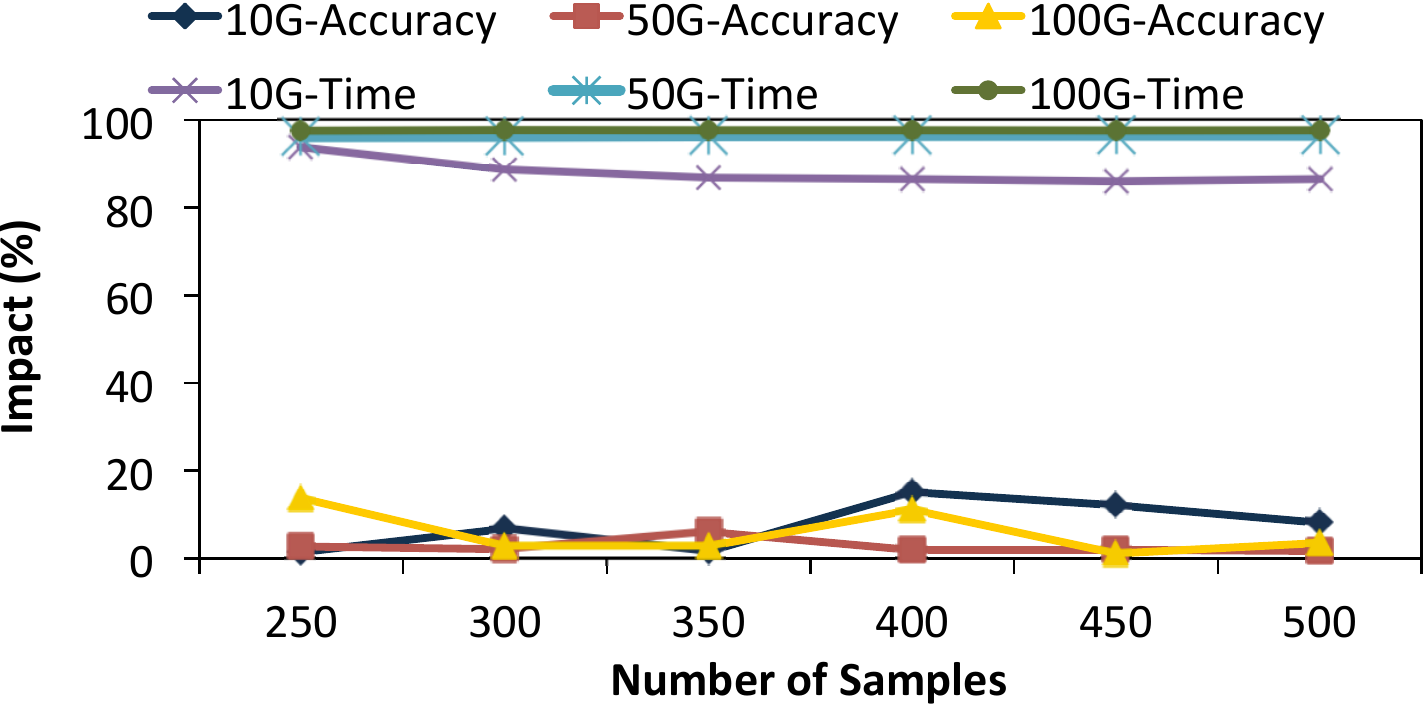}
\label{f:fmeasure_sampledb}}
\subfigure[\scriptsize{Time  improvement of sampled datasets and increasing number of areas  ($>$70\% accuracy, large areas).}]{
\includegraphics[totalheight=1.03in, angle=0]{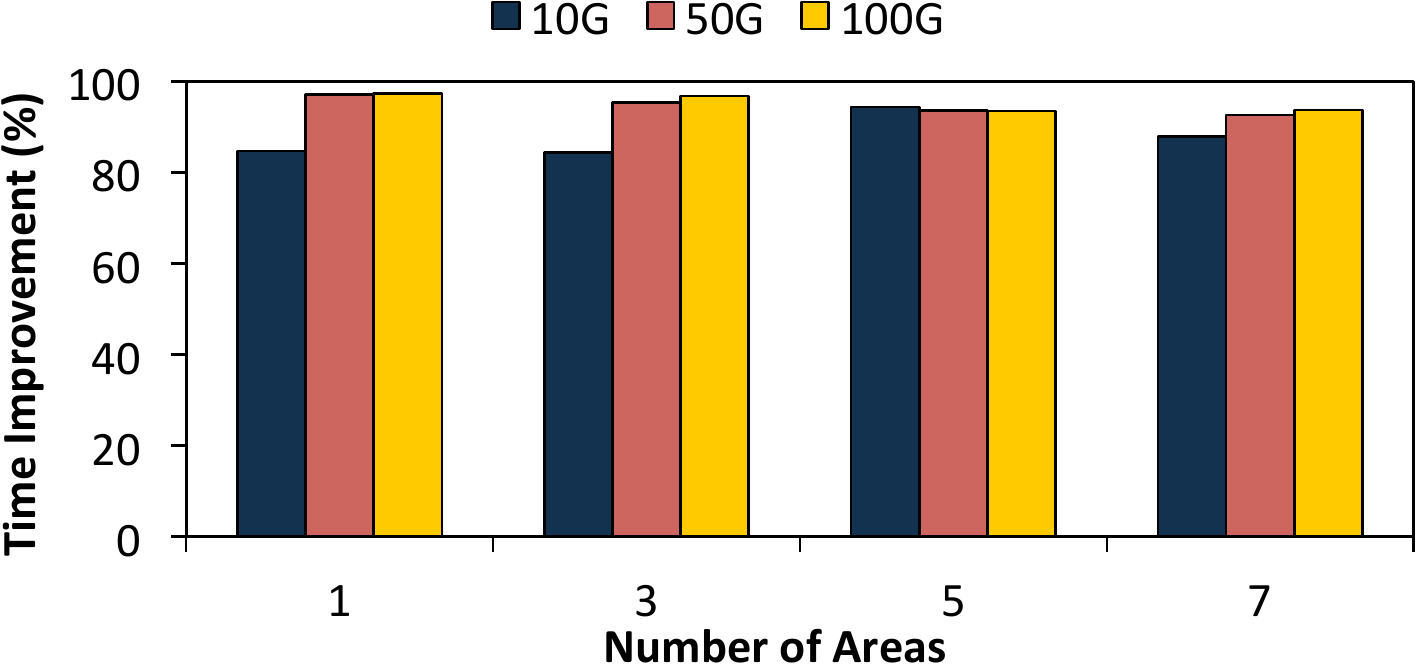}
\label{f:timeareas_sampledb}}
\vspace{-3mm}
\caption{\small{Figure (a) shows AIDE's effectiveness on big data sets and Figures (b-c) show the impact of our exploration space reduction (b-c).}} 
\label{f:scalability}
\vspace{-4mm}
\end{figure*}

{\bf Database Size}  Figure~\ref{f:fmeasure_dbsize} shows AIDE's accuracy with a given number of labeled samples for dataset sizes of 10GB, 50GB and 100GB. Our target queries have one large relevant area and the average number of relevant objects increases as we increase the size of the dataset (our target query returns in average 26,817 relevant objects in the 10GB, 120,136 objects in the 50GB and 238,898 objects in the 100GB database). AIDE predicts these objects in all datasets with high accuracy without increasing the user's effort. \emph{We conclude that the size of the database does not affect our effectiveness.} AIDE consistently achieves high accuracy of more than 80\% on big data sets with only a few hundred samples (e.g., 400 samples).  These results were consistent even for more complex queries with multiple relevant areas. 

{\bf Exploration Space Reduction} Applying our techniques to larger datasets increases  the time overhead since our sampling queries have higher response times.  One optimization is to execute our exploration on a sampled database (Section~\ref{s:sparereduction}). In this experiment,\cut{we applied our exploration on sampled databases with size 10\% of the size of the original ones and we  evaluated the impact on our effectiveness and efficiency. Specifically, } we sampled  datasets of  10GB, 50GB, 100GB and generated the  10\% sampled datasets of 1GB, 5GB and 10GB, respectively. Figure~\ref{f:fmeasure_sampledb} shows the absolute difference of the final accuracy (\emph{10GB-Accuracy, 50GB-Accuracy, 100GB-Accuracy}) when AIDE is applied on the sampled and on the total datasets.  The average difference is no more than 7.15\% for the 10GB, 2.72\% for the 50GB and 5.85\% for the 100GB data set. In the same figure we  also show the improvement of the system execution time  (\emph{{10GB-Time, 50GB-Time, 100GB-Time}}). For 10GB (and a sampled dataset of 1GB) this time is reduced  by 88\% in average, while for the larger datasets of 50GB and 100GB it is reduced by 96\%-97\%.

Figure~\ref{f:timeareas_sampledb}  shows the improvement of the system execution time when AIDE runs over the sampled data sets and we  increase the number of  relevant areas. Here, we measure the  improvement of the system execution time when we reach an accuracy higher than 70\% . The average time per iteration is 2.8 seconds for the 10GB, 37.7 for the 50GB and 111  for the 100GB database. By operating on the sampled datasets we improved our time by more than 84\% while our average improvement for each query type was more than 91\%. Our improved iteration time is 0.37 second for the 10GB, 2.14 seconds for the 50GB and 5.3 seconds for the 100GB dataset, in average. {The average number of iterations is 37 and hence  AIDE offers a total execution time of 13secs for the 10GB, 1.3mins for the 50GB and 3.2mins for the 50GB dataset while the user wait time is less than 3secs per iteration in average.  \emph{Hence, AIDE can scale to big datasets by applying its  techniques  on sampled datasets. This incurs  very low impact on the accuracy while it significantly improves the system execution time.  }}

%

{\bf Exploration Space Dimensionality} Figure~\ref{f:5D_fmeasure} shows the number of samples to reach an accuracy  greater than 70\% as we increase the complexity of our queries (the number of relevant areas) and the size of our exploration space from 2-dimensional to 5-dimensional. These results are on large size areas and on the sampled datasets. Our target queries have conjunctions on two attributes and the main challenge for AIDE is to identify in the 3D, 4D and 5D spaces only the two relevant attributes.\emph{AIDE correctly identifies the irrelevant attributes and eliminates them from the decision tree classifier and hence from the final output query. Furthermore, although the exploration over more dimensions requires naturally more samples to reach an acceptable accuracy, the number of samples only increases by a small percentage} (the 3D space and 4D space require in average 13\% more tuples than the 2D space and the 5D space requires 32\% more tuples than the 2D space) and they remain within the range of 100's even for the complex cases of 7 areas and 5-dimensional exploration space. Figure~\ref{f:5D_time} shows that even for the very complex case of seven (7) relevant areas the time overhead is always less than 4.5 seconds, while for the less complex queries of 1 area the time drops below 1 second. {These results reveal a small increase in the user's wait time as we add more dimensions (each new dimension adds in average 0.7 seconds overhead to the previous one) but always within acceptable bounds.  }

\subsection{User Study Evaluation}\label{s:userstudy}


Our user study used  the AuctionMark dataset~\cite{AuctionMark} that includes  information on auction items and their bids. 
We chose this ``intuitive" dataset, as opposed to the  SDSS dataset, because the user study requires identifying a significant number of users with sufficient   understanding of the domain. 
Thus, AuctionMark meets the requirement: we were able to identify a group of computer science graduate students with SQL experiences and designed their exploration task to be ``identifying auction items that are good deals''.
Note that the exploration task  should not be trivial, i.e.,  users should not have an upfront understanding of the exact selection predicates that would collect all relevant objects. 

The exploration data set had a size of 1.77GB and it was derived from the ITEM table of AuctionMark benchmark. It included seven attributes: initial price, current price, number of bids, number of comments, number of dates an item is in an auction, the difference between the initial and current item price, and the days until the auction is closed for that item. Each user explored the data set ``manually", i.e., iteratively formulating exploratory queries and reviewing their results until they obtained a query, $Q$, that satisfied their interests. 
We then took $Q$ as the true interest of a user and used it to simulate the user labeling results in  AIDE. We measured how well AIDE can predict $Q$. 

The results demonstrated that AIDE was able to reduce the user's reviewing effort by 66\% in average (\textit{Reviewing savings} column in Table~\ref{tab:userstudytable}). Furthermore, with the manual exploration users were shown 100s of thousands  objects in total (\textit{Manual returned objects}) while AIDE shows them only a few hundred strategically selected samples. Furthermore, with the manual exploration our users needed about an hour to complete their task (\emph{Manual time}). Assuming that the most of this time was spent on tuple reviewing, we calculated the average tuple reviewing for each user. This varied significantly across users (3secs - 26secs).  Using this time we estimated the total exploration time needed by AIDE including the reviewing effort (\emph{AIDE time}). AIDE was able to reduce the exploration time 47\% in average. We believe these time savings will be even more pronounced for more complex exploration tasks (e.g., in astronomical or medical domains) where examining the relevance of an object requires significant time.

Our user study revealed that five out of the seven users used only two attributes to characterize their interests \cut{while the rest needed three, four and five attributes}. Similarly to our SDSS workload, the most common type of query was  conjunctive queries that selected a single relevant area.  {Our exploration domain was highly skewed and all our relevant areas were on  dense regions.}  These characteristics indicate that our  micro-benchmark on the SDSS dataset was representative of common exploration tasks while it also covered highly more complex cases, i.e.,  small relevant areas and disjunctive queries selecting multiple areas. 


\begin{table}[t]
\centering
\scriptsize
\begin{tabular}{ |c|r|r|r|r|r|r|}
\hline
\textbf{User} & \textbf{Manual:} & \textbf{Manual:} &  \textbf{AIDE:} & \textbf{Reviewing}   & \textbf{Manual:} &\textbf{AIDE:}  \\
 & \textbf{returned} &  \textbf{reviewed}  & \textbf{reviewed}  & \textbf{savings}  & \textbf{time} & \textbf{time} \\
 &  \textbf{objects} &   \textbf{objects} &  \textbf{objects} &  (\%) & (min) & (min) \\
 \hline

  1    & 253,461 &312 & 204.9 & \textbf{34.3\%} & 60 & 39.7\\
  2 & 656,880  & 160 & 82.4 & \textbf{48.5\%} & 70 &  36.3\\
  3 & 933,500 & 1240 & 157 & \textbf{87.3\%} & 60 &  7.9\\
  4 & 180,907 & 600 & 319 & \textbf{46.8\%} & 50 &  28.2\\ \hline
 5 &  2,446,180 & 650  & 288.5 & \textbf{55.6\%} & 60 &  27.5\\ \hline
 6   & 1,467,708 & 750 & 334.5 & \textbf{55.3\%} & 75 & 33.8\\ \hline
 7 & 567,894  & 1064  & 288.4 & \textbf{72.8\%} & 90 &  24.8\\ 
\hline
\end{tabular}
\vspace {-3mm}
\caption{User study results.}
\vspace {-4mm}
  \label{tab:userstudytable}
\end{table}

\section{Related Work}\label{s:related}

{\bf Query by Example} Related work  on ``Query-By-Example'' (QBE) we originally proposed in~\cite{ zloof75}. Most recent work includes querying knowledge graphs by example tuples~\cite{mottin14}, formulating join queries based on example output tuples~\cite{yanyan14} and inferring  user queries by asking for feedback on database tuples~\cite{haoli15, bonifati_edbt14}. Finally, in~\cite{abouzied13} they learn user queries based on given value assignments used in the intended query. 
These systems  provide alternative front-end  query interfaces that assist the user formulate her query and do not attempt to understand user interests nor retrieve ``similar'' data objects which is AIDE's focus.

{\bf Data Exploration}  Numerous recent research efforts focus on data exploration. The vision for automatic, interactive navigation in databases  was first discussed in~\cite{ugur13} and later on in~\cite{wasay15}.  \cut{AstroShelf~\cite{astroshelf2012_ss} allows users to collaboratively annotate and explore sky objects while }YMALDB~\cite{ymaldb2013} supports data exploration by recommending to the user data similar to her query results. DICE~\cite{dice2014_ss} supports exploration of data cubes using faceted search and in~\cite{manas15}  they propose a new ``drill-down'' operator for exploring and summarizing groups of tuples. SciBORQ~\cite{sciborq11} relies on hierarchical database samples to support scientific exploration queries within strict query execution times.  \cut{Blink~\cite{sameer12} relies on run-time sample selection to provide real-time answers with statistical error guarantees.} Idreos et al.~\cite{kersten11} envision a system for interactive data processing tasks aiming to reduce the time spent on data analysis. In~\cite{sellam13} interactively explores the space based on statistical properties of the data and provides query suggestions for further exploration while in~\cite{lilong15} they propose a technique for providing feedback during the query specification and eventually guiding the user towards her intended query. In~\cite{alex14} users rely on prefetching and incremental online processing  to offer interactive exploration times for window-based queries. 
SearchLight~\cite{alex15}  offers fast searching, mining and exploration of multidimensional data based on constraint programming. All the above systems are different than AIDE: we rely on the user's feedback on data samples to predict the user's data interests and we focus on identifying strategic sampling areas that allow for accurate predictions.


{\bf Query Relaxation} Query relaxation  techniques have also been proposed for supporting exploration in databases~\cite{chaudhuri90}. In ~\cite{mishra09, koudas06_ss} they refine SQL queries to satisfy cardinality constraints on the query result.  In~\cite{kadlag04} they rely on multi-dimensional histograms and distance metrics for range queries for accurate query size estimation. These solutions are orthogonal to our problem; they focus on adjusting the query parameters to reach a cardinality goal and therefore cannot characterize user interests.


{\bf Active Learning} The active learning community has  proposed solutions that maximize the learning outcome while minimizing the number of samples labeled by the user~\cite{roy2001, sarawagi02}. However, these techniques assume either small datasets or negligible sample extraction costs which is not a valid assumption when datasets  span 100s of GBs and  interactive performance is expected. Relevance feedback have been studied for image retrieval~\cite{Panda:2006:ALV}, document ranking~\cite{ruthven03}, information extraction and segmentation~\cite{settles2008} and word disambiguation~\cite{zhu2007}. All these solutions are designed for  specific data types (images or text) and do not optimize for efficient {sample acquisition} and {data space exploration}.

{\bf Collaborative and Interactive Systems}
In~\cite{nodira} a collaborative system  is proposed to facilitate formulation of SQL queries based on past queries and in~\cite{gloria09} they use collaborative filtering to provide  query recommendations. However, both these systems do not predict ``similar'' data object.  In~\cite{sadikov10} they cluster  related queries as a means of understanding the  intents of a given user query. The focus is on web searches and not structured databases.

\section{conclusions}\label{s:conclusions}

{Interactive Data Exploration (IDE) is a key ingredient of a diverse set of discovery-oriented applications, including ones from scientific computing and evidence-based medicine. In these applications,  data discovery is a highly ad hoc interactive process where users execute numerous exploration queries using varying predicates aiming to balance the trade-off between collecting all relevant information and reducing the size of returned data. Therefore, there is a strong need to support these human-in-the-loop applications by assisting their navigation in the data space.}

In this paper, we introduce AIDE, an \emph{Automatic Interactive Data Exploration} system, that iteratively steers the user towards interesting data areas and ``predicts'' a query that retrieves her objects of interest. Our approach leverages relevance feedback on database samples to model user interests and strategically collects more samples to refine the model while  minimizing the user effort. AIDE integrates machine learning and data management techniques to provide effective data exploration results (matching the user's interests with high accuracy) as well as high interactive performance. It delivers highly accurate query predictions for very common conjunctive queries with very small user effort while, given a reasonable number of samples, it can predict with high accuracy complex conjunctive queries. Furthermore, it provides interactive performance by limiting the  user wait time per iteration to less than a few seconds in average. Our user study indicates that AIDE is a practical exploration framework as it significantly reduces the user effort and  the total exploration time compared with the current state-of-the-art approach of manual exploration.


%
%
%

\bibliographystyle{IEEEtran}
\bibliography{qsteering}

\vspace {-14mm}

\begin{IEEEbiography}[{\includegraphics[width=1in,height=1.25in,clip,keepaspectratio]{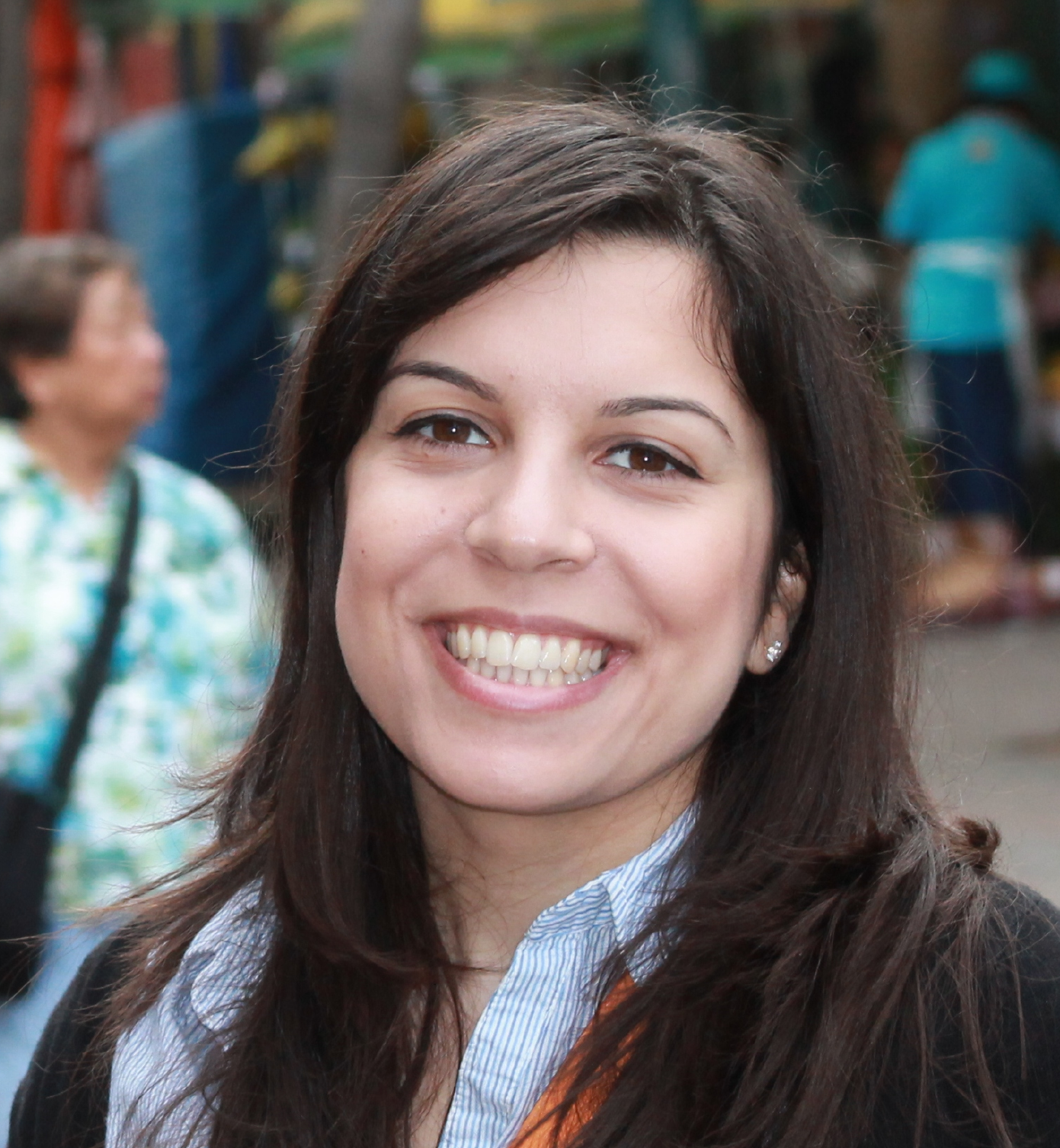}}]{Kyriaki Dimitriadou}
Kyriaki Dimitriadou is a PhD student in Computer Science at Brandeis University. She holds an MA in Computer Science from Brandeis and a BA in Applied Informatics from the University of Macedonia, Greece. Her research interests are in database systems with a focus on  data exploration. 
\end{IEEEbiography}

\vspace {-20mm}

\begin{IEEEbiography}[{\includegraphics[width=1in,height=1.25in,clip,keepaspectratio]{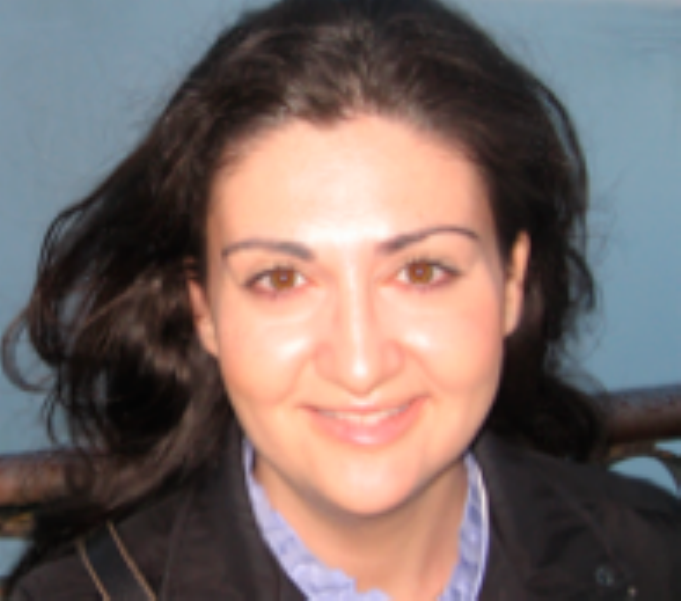}}]{Olga Papaemmanouil}
Olga Papaemmanouil is an Assistant Professor of Computer Science at Brandeis University since 2009. She received her undergraduate degree in Computer Engineering and Informatics from the University of Patras, Greece,, a M.Sc. in Information Systems from the Athens University of Economics and Business and completed her PhD at Brown University in 2008. Her research interests are in data management and distributed systems with  focus on data streams, query performance, cloud databases and  data exploration. She is the recipient of an NSF CAREER Award (2013) and a Paris Kanellakis Fellow (2002).
\end{IEEEbiography}

\vspace {-14mm}

\begin{IEEEbiography}[{\includegraphics[width=1in,height=1.25in,clip,keepaspectratio]{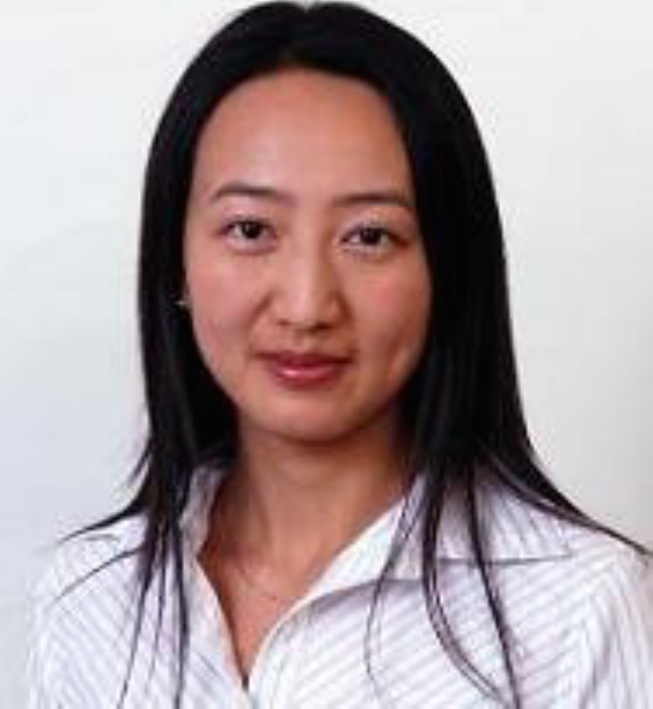}}]{Yanlei Diao}
received the Bachelors degree in computer science from Fudan University in China in 1998, the M.Phil degree from the Hong Kong University of Science and Technology in 2000, and the Ph.D. degree in computer science from the University of California, Berkeley in 2005. She is an Assistant Professor in the Department of Computer Science at the University of Massachusetts. Her
research interests are in information architectures and data management systems, with a focus on data
streams, sensor data management, uncertain data management, large-scale data analysis, and flash memory databases. Dr. Diao
has been the recipient of the NSF CAREER Award, the
finalist for the Microsoft Research New Faculty Fellowship, and the recipient
of the IBM Scalable Innovation Faculty Award. 
Her PhD dissertation won the 2006 ACM-SIGMOD Dissertation Award
Honorable Mention.

\end{IEEEbiography}

\end{sloppypar}
\end{document}